\documentclass[pre,10pt,twocolumn,notitlepage,floatfix,footinbib,superscriptaddress,showpacs,showkeys,aps]{revtex4-1}

\usepackage[T1]{fontenc}
\usepackage[latin9]{inputenc}

\usepackage{float}
\usepackage{amsmath,amssymb}
\usepackage{graphicx}

\usepackage{enumitem}

\usepackage[colorlinks]{hyperref}
\usepackage{xcolor}

\usepackage{natbib}

\graphicspath{{./figures/}{.}}

\begin{document}

\title{Many-body quantum dynamics of initially trapped systems due to a Stark potential ---
  thermalization vs.\ Bloch oscillations}

\author{Pedro Ribeiro}\email{ribeiro.pedro@gmail.com}
\affiliation{CeFEMA, Instituto Superior T\'{e}cnico, Universidade de Lisboa, Av. Rovisco Pais, 1049-001 Lisboa, Portugal}

\affiliation{Beijing Computational Science Research Center, Beijing 100193, China}

\affiliation{Max Planck Institute for the Physics of Complex Systems, Nothnitzer Str. 38, 01187 Dresden, Germany}

\author{Achilleas Lazarides}
\affiliation{Max Planck Institute for the Physics of Complex Systems, Nothnitzer Str. 38, 01187
  Dresden, Germany}

\affiliation{Interdisciplinary Centre for Mathematical Modelling and Department of Mathematical Sciences, Loughborough
University, Loughborough, Leicestershire LE11 3TU, United Kingdom}

\author{Masudul Haque}

\affiliation{Max Planck Institute for the Physics of Complex Systems, Nothnitzer Str. 38, 01187 Dresden, Germany}

\affiliation{Department of Theoretical Physics, Maynooth University, Co.~Kildare, Ireland}

\begin{abstract}

We analyze the dynamics of an initially trapped cloud of interacting quantum particles on a lattice
under a linear (Stark) potential.  We reveal a dichotomy: initially trapped interacting systems
possess features typical of both many-body-localized and self-thermalizing systems.  We consider
both fermions ($t$-$V$ model) and bosons (Bose-Hubbard model).  For the zero and infinite
interaction limits, both systems are integrable: we provide analytic solutions in terms of the
moments of the initial cloud shape, and clarify how the recurrent dynamics (many-body Bloch
oscillations) depends on the initial state.
Away from the integrable systems, we identify and explain the time scale at which Bloch oscillations
decohere.

\end{abstract}

\maketitle

\begin{minipage}[t]{1\columnwidth}%
\global\long\def\ket#1{\left| #1\right\rangle }

\global\long\def\bra#1{\left\langle #1 \right|}

\global\long\def\kket#1{\left\Vert #1\right\rangle }

\global\long\def\bbra#1{\left\langle #1\right\Vert }

\global\long\def\braket#1#2{\left\langle #1\right. \left| #2 \right\rangle }

\global\long\def\bbrakket#1#2{\left\langle #1\right. \left\Vert #2\right\rangle }

\global\long\def\av#1{\left\langle #1 \right\rangle }

\global\long\def\tr{\text{Tr}}

\global\long\def\pd{\partial}

\global\long\def\im{\text{Im}}

\global\long\def\re{\text{Re}}

\global\long\def\sgn{\text{sgn}}

\global\long\def\Det{\text{Det}}

\global\long\def\abs#1{\left|#1\right|}

\global\long\def\up{\uparrow}

\global\long\def\down{\downarrow}

\end{minipage}

% WARNING: THE PAPER USES (J/2) AS THE HOPPING AMPLITUDE, NOT J.  THIS MEANS THAT THE E VALUES IN MASUD'S FIGURES SHOULD BE HALVED FROM THE NOTATION USED IN MASUD'S CALCULATIONS.

\emph{Introduction---} The historical focus of many-body quantum physics has been the low-energy
parts of the many-body spectrum.
%
%% This is well-justified in solid-state systems which are usually in contact with a thermal bath and
%% typically relax fast to low-energy sectors.
%
In recent years, the perspective has changed, largely due to experiments with cold atoms
\cite{BlochDalibardZwerger_RMP2008, Morsch_Oberthaler_RMP2006,
  Weiss_QuantumNewtonsCradle_Nature2006, GreinerBloch_Nature2002dynamics}, which have inspired the
study of non-equilibrium situations in \emph{isolated} quantum systems
\cite{PolkovnikovRigol_AdvPhys2016, EisertGogolin_NatPhys2015, PolkovnikovSenguptaSilva_RMP2011,
  Dziarmaga2010}.  In an isolated situation, energy conservation ensures that a system with an
initially high energy will never explore the low-energy parts of the spectrum.  The quantum dynamics
of isolated systems poses new challenging questions, such as
%
% whether observables equilibrate at long times to values predicted by a thermal ensemble
%
% the question of \emph{thermalization} 
%
whether observables \emph{thermalize} \cite{PolkovnikovRigol_AdvPhys2016, EisertGogolin_NatPhys2015,
  PolkovnikovSenguptaSilva_RMP2011, Rigol2008}.

A well-known example for which isolation leads to drastically different dynamics is the phenomenon
of Bloch oscillations \cite{Bloch1929,Zener1934}.  Particles in a tight-binding lattice subject to a
linear potential, e.g., due to gravity or an electric field, do not accelerate toward lower
potentials, but rather undergo local oscillations.  For a single particle, the shape and/or position
of the particle wavefunction oscillates, perfectly periodically
\cite{Bloch1929,Zener1934,Hartmann2004,Thommen_2004a}.  Long after its prediction, Bloch
oscillations were observed in semiconductor super-lattices \cite{Waschke1993,Lyssenko1997}, in cold
atoms \cite{BenDahan1996,NiuRaizen_PRL1996,Raizen1997,AndersonKasevich_Science1998}, and in periodic
photonic structures \cite{Morandotti-etal_optical_PRL1999,Pertsch1999, Sapienza2003}.
In cold atom experiments, Bloch oscillations have by now been observed many times
\cite{BenDahan1996,NiuRaizen_PRL1996,Raizen1997,AndersonKasevich_Science1998,
  Morsch2001,Morsch_Arimondo_PRA2002, Roati2004, Battesti2004, 
  Tino_BlochOscilExp_PRL2006,  Morsch_Oberthaler_RMP2006, Biraben_BlochOscilExpt_PRA2006, 
 Drenkelforth2008, Naegerl_Bloch_dephasing_PRL2008,
  Naegerl_superbloch_PRL2010, Poli_et_al_precision_gravity_PRL2011, Tarruell2012, Tino_BlochOscilExpt_PRA2012, 
  electric_quantumwalks_PRL2013, Meinert_PRL2014, Greiner_quantumwalk_Science2015,
  Hemmerich_NJP2016, Hemmerich_PRL2017, Weld_positionspaceBloch_PRL2018},
and are used widely as a measurement tool, e.g., for metrological applications \cite{Battesti2004,
  Biraben_BlochOscilExpt_PRA2006, Poli_et_al_precision_gravity_PRL2011, Tino_BlochOscilExpt_PRA2012}
to detect Dirac points in optical lattices \cite{Tarruell2012}, etc.  Some experiments have also
explored the effect of inter-particle interactions on Bloch oscillations
\cite{Morsch2001,Roati2004,Naegerl_Bloch_dephasing_PRL2008, Drenkelforth2008,
  Naegerl_superbloch_PRL2010, Meinert_PRL2014}.
Theoretical treatments of Bloch oscillations have addressed a variety of single-particle situations
\cite{Kolovsky2002,Hartmann2004,Thommen_2004a,Breid2006,Breid2007,Thommen_2004b,
  Durst_aperiodic_BlochOscil_PRA2010, Longhi_BlochOscil_defects_PRB2010, Collura2012,
  KolovskyBulgakov_PRA2013, KhomerikiFlach_BlochOscil_flatband_PRL2016,
  KartashovEA_BlochOscil_SOC_PRL2016, Longhi_accelerated_AiryBloch_IJMPB2016,
  ZhengFengYang_fluxladder_PRA2017}, interacting few-particle systems
\cite{DiasEA_2pcle_BlochOscil_PRB2007, Krimer2009, DiasEA_2pcle_BlochOscil_PRB2010,
  Longhi_2anyon_BlochOscil_PRB2012, Zakrzewski_2bosons_PRA2017}, and interacting many-body systems
\cite{Buchleitner2003,Kolovsky2003,Kolovsky2004, Breid2007, Schulte_Lewenstein_LSantos_PRA2008,
  Salerno_Blochoscill_nonlin_PRL2008, Kolovsky2009,Kolovsky2010,Mierzejewski2010,
  Durst_aperiodic_BlochOscil_PRA2010, Eckstein2010b, Eckstein2011a,Witthaut2011, Rubbo2011,Cai2011,
  Mandt2011, GaulMueller_timedepnonlinearity_PRA2011, Longhi2012, Carrasquilla2013,
  Mandt_Blochdamping_BoltzmannEq_PRA2014, Mahmud2014, Driben_nonlin_Blochosc_SciRep2017}.
Interactions have been treated both in mean-field (e.g., Gross-Pitaevskii) regimes
\cite{Kolovsky2004, Breid2007, Salerno_Blochoscill_nonlin_PRL2008,
  Schulte_Lewenstein_LSantos_PRA2008, Kolovsky2009, Durst_aperiodic_BlochOscil_PRA2010,
  Kolovsky2010, Witthaut2011, GaulMueller_timedepnonlinearity_PRA2011,
  Driben_nonlin_Blochosc_SciRep2017} and beyond the mean-field regime
\cite{Buchleitner2003,Mierzejewski2010, Eckstein2010b, Eckstein2011a,
  Rubbo2011,Carrasquilla2013,Mahmud2014}.

Recent experiments \cite{Meinert_PRL2014} have found, by tuning bosonic on-site repulsion, the
collapse and revival of the oscillation of the cloud position, with the revival period proportional
to interaction strength.  In addition, sufficiently far from the non-interacting point, the atom
cloud was found to have `chaotic' behavior leading to rapid relaxation.

In this work, we address the real-time dynamics of an initially trapped interacting lattice system
subject to a linear potential.  We present a comprehensive study for two representative systems
(featuring bosons and fermions), for all interaction regimes.  At zero or infinite interaction, each
model becomes integrable (can be mapped to free particles).  For intermediate interactions, we have
an example of many-body localization without disorder \cite{Gavish2005, Yao2014, Schiulaz2015,
  Antipov2016, Smith.2017, Mondaini2017, Schulz.2018}, where nevertheless a version of
thermalization is valid when we focus on the part of the Hilbert space spanned by states in which
particles are confined within a connected spatial region, i.e., the subspace explored by initially
trapped systems.  We show that the dynamics within such a subspace is thermalizing.

%if we focus on the part of the Hilbert space in which the particles are
%trapped within a connected part of the lattice, the dynamics within this subspace is thermalizing.

At the ``free'' points, there is perfectly periodic behavior.  We provide a series of exact
analytical results for the cloud dynamics in these cases.  For strongly interacting (hardcore)
bosons, we show dynamical generation (and periodic disappearance!) of fragmented condensation of an
initial un-condensed cloud.  At strong (weak) initial trapping, the dynamics consists primarily of
width (position) oscillations.  At intermediate trapping, the skewness undergoes unusual dynamics
during every period, of which we do not know of an analog in the literature.
Near the integrable points, we show and explain beating behavior of the cloud dynamics, with linear
dependences on the integrability-breaking parameter.  This explains and generalizes the experimental
observation of \cite{Meinert_PRL2014}.

\emph{Models---} We consider $N_{p}$ particles on an infinite lattice subjected to a tilt potential.
The total Hamiltonian
\begin{equation}
H=  \mathcal{T} + \mathcal{E} + \mathcal{V}  \label{eq:H},
\end{equation}
consists of a kinetic term $\mathcal{T} = -J/2\sum_{j}
\left(a_{j}^{\dagger}a_{j+1}+\text{h.c.}\right)$, with $a_{j}^{\dagger}$ the creation operator of a
particle in site $j$ and $J$ the hopping amplitude; a potential term $ \mathcal{E} =E\sum_{j}j\,
a_{j}^{\dagger}a_{j} $ due to a constant tilt strength $E$; and an interaction term $ \mathcal{V} $.
We consider two families of models: the Bose-Hubbard model (BHM) for which $
\mathcal{V} = U/2 \sum_j b_{j}^{\dagger}b_{j}^{\dagger}b_{j}b_{j} $, in which case the particles are
bosons $a_{j}=b_{j}$; and the $t$-$V$ model (Ft-VM) with $\mathcal{V} = V \sum_j
c_{j}^{\dagger}c_{j+1}^{\dagger}c_{j+1}c_{j} $, featuring interacting spinless fermions
$a_{j}=c_{j}$.

We will mostly take the initial state $\ket{\Phi_0}$ to be the ground state of the non-tilted system
in the presence of a harmonic potential ($H_{0}= \mathcal{T} + \mathcal{V} +W\sum_{j}j^2\,
a_{j}^{\dagger}a_{j}\label{eq:H_i}$), parametrized by the dimensionless constant
$\tilde{\rho}=N_{p}\sqrt{W/J}$ \cite{RigolMuramatsu_QMC_PRA2004,
  RigolMuramatsu_confinementcontrol_PRA2004}.  The initial condition can be varied form an extended
Gaussian-like cloud (small $\tilde{\rho}$) to a highly packed state at large $\tilde{\rho}$.  We
also consider initial states which are product states, e.g., of the form $\ket{\Phi_0} =
a_{i+1}^\dagger a_{i+2}^\dagger\ldots a_{i+{N_p}}^\dagger \ket{0}$.  For bosons at $U=\infty$ and
for fermions at all $V\neq\infty$, the ground state has this form at large $\tilde{\rho}$.

In addition to $U,V=0$, in both strong interacting regimes ($U,V\to \infty$), the dynamics is that
of a set of non-interacting particles.  For $U\to\infty$ (BHM), double occupancy is kinematically
forbidden and the finite energy Hilbert space reduces to that of hard core bosons. In this limit the
BHM maps to the Ft-VM with $V=0$ via a Jordan-Wigner (JW) transformation.  The spectrum of the Ft-VM
with $V\to\infty$ and $L$ sites can also be shown to map onto that of a Ft-VM with $L-N_p$ sites and
$V=0$ \cite{supmat}. In all these (effectively) non-interacting cases, the spectrum of the tilted
Hamiltonian consists of equally spaced highly degenerate levels, with spacing $E$.  This yields
periodic evolution, with period $T=2\pi/E$, for any initial state.  In fact, exact analytical
solutions can be found for the many-body evolution \cite{supmat}. Away from these `free' cases the
dynamics is non-integrable: either because the non-tilted model is already so (BHM); or because a
finite tilt breaks the integrability present in the $E=0$ case (Ft-VM).

\begin{figure}
\centering
\includegraphics[width=1\columnwidth]{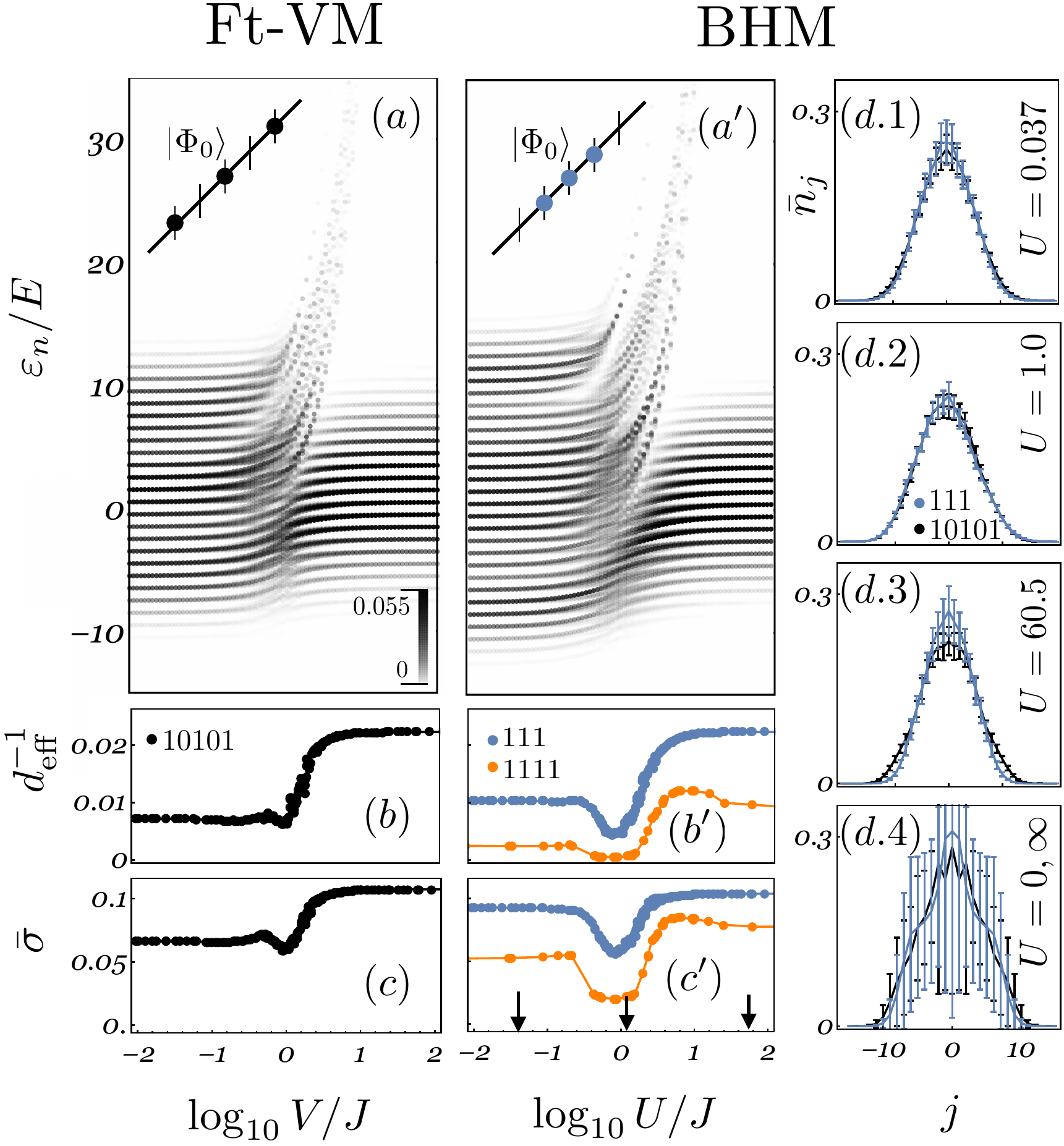}
\caption{\label{fig:sigma_2} $(a, a')$ Spectra of the Ft-VM and of the BHM as a function of the
  interaction strength. The color coding corresponds to the overlap-squared, $\abs{c_\alpha}^2$. of
  the initial state, which is $\ket{...0101010...}$ for the Ft-VM and $\ket{...01110...}$ for
  BHM. 
$(b, b')$ Inverse of effective dimension, $d_\text{eff}$.  
$(c, c')$ Strength $\bar{\sigma}$ of temporal fluctuations around the asymptotic long time density
  average, for $N_p=3$ and $4$.
$(d.1\text{-}4)$ Asymptotic average of the density profile, $\bar n_j$, for two initial
  states with similar energies.  The error bars depict the fluctuations $\bar \sigma_j$ at each
  site.  3 values of $U$ [marked by arrows in $(c')$] and $U=0,\infty$ are used.
}
\end{figure}

\emph{Long time behavior and thermalization---} Figs.\ \ref{fig:sigma_2}($a$,$a'$) show the
eigenenergies $\varepsilon_\alpha$ of $H$, corresponding to eigenvectors $\ket \alpha$, as a
function of the interaction strength, color-coded with $\abs{c_\alpha}^2$, with $c_\alpha =
\braket{\alpha}{\Phi_0}$ the overlap amplitude with the initial state.  Only some eigenstates have a
non-negligible overlap with the initial state; the other eigenenergies are not visible.  For fixed
$N_p$, Increasing the chain length $L$ (with fixed $N_p$) increases the Hilbert space dimension
polynomially, rendering the spectrum dense at $L\to\infty $, but leaves
Figs.\ \ref{fig:sigma_2}($a$,$a'$) invariant.  Density profiles of the many-body eigenstates which
have non-negligible $\abs{c_\alpha}^2$ are exponentially localized within a length proportional to
$1/E$. Therefore the dynamics of an initially confined cloud of atoms is always localized.  This can
be traced to the fact that a cloud of atoms in an infinite system is always in the dilute density
regime; as interactions are short-range, if the cloud expands too much the particles cease to
interact with each-other. The exponential localization of the many-body eigenstates, and
consequently of the dynamics, is thus ensured by the exponential localization of the single particle
eigenstates \cite{Bloch1929,Zener1934}.

The effective dimensionality of the Hilbert space spanned by the initial state is $d_\text{eff} =
\left( \sum_\alpha \abs{c_\alpha}^4 \right)^{-1}$ \cite{Popescu2006}, shown in
Figs.\ \ref{fig:sigma_2}($b$,$b'$). This quantity is larger for intermediate interactions than near
the `free' points (small or large $U$, $V$).  $d_\text{eff}$ decreases algebraically with $E$ and
increases algebraically with the number of particles $N_p$.

We now analyze the long time asymptotic behavior of the cloud dynamics in light of these spectral properties. 
We define the time averaged density $\bar n_j = \lim_{T\to\infty} T^{-1} \int_0^T dt\, n_j(t) $, with $n_j(t)=\av{a_{j}^{\dagger}(t)a_{j}(t)}$ the site occupancy, and  $\bar \sigma_j ^2 =  \lim_{T\to\infty} T^{-1} \int_0^T dt [ n_j(t) -\bar n_j]^2  $, which quantifies the temporal deviations around the average.  For a system with a non-degenerate spectrum these quantities are given by their diagonal ensemble \cite{Rigol2008} values $\bar n_j=\sum_\alpha \abs{c_\alpha}^2 \bra{\alpha }n_i\ket{\alpha}$ and $\bar \sigma_j ^2 = \sum_{\alpha\neq\alpha'} \abs{c_\alpha}^2 \abs{c_ {\alpha'}}^2 \abs{\bra{\alpha }n_i\ket{\alpha'}}^2$. 
Some representative density profiles and $\bar \sigma = (\sum_j \sigma_j^2)^{1/2}$ for different values of $U$ and $V$ are depicted in Figs.\ \ref{fig:sigma_2}($c$,$c'$). 

For systems fulfilling the so called eigenstate thermalisation hypothesis (ETH) \cite{Deutsch1991,
  Srednicki1994,Rigol2008} the temporal fluctuations of local observables are strongly suppressed,
decreasing exponentially with system size.  In contrast, for integrable models, ETH does not hold:
The decrease is merely polynomial.  In the present case the system does not fulfill ETH trivially -
there are an infinite number of eigenstates with the same energy but a vanishing overlap with the
initial state. Moreover, as all eigenstates are localized throughout the spectrum, the system
behaves as a many-body-localized (MBL) one.

Nonetheless, away from the `free' points, equilibration may still arise for a sufficiently
large $N_p$, i.e. large $d_\text{eff}$, in the sense that different trapped initial states with
roughly the same energy yield the same $\bar n_i$ profile and that long time deviations from the
average are suppressed $\bar \sigma \propto 1/d_\text{eff}$.  A comparison between
Figs.~\ref{fig:sigma_2}$(b,b')$ and $(c,c')$ shows that $d_\text{eff}^{-1}$ and $\bar \sigma$ are
qualitatively similar and that the values of $\bar \sigma$ substantially decrease with the number of
particles in the cloud. This supports an equilibration scenario for both fermionic and bosonic
systems away from $U,V=0$ and $U,V=\infty$. At these special values the system becomes integrable
and the limits $U,V\to0,\infty$ and $t\to\infty$ do not commute.  At these points $\bar\sigma$ is
much larger and decreases much slower with particle number.

\begin{figure}
\begin{centering}
\begin{tabular}{c}
\includegraphics[width=1\columnwidth]{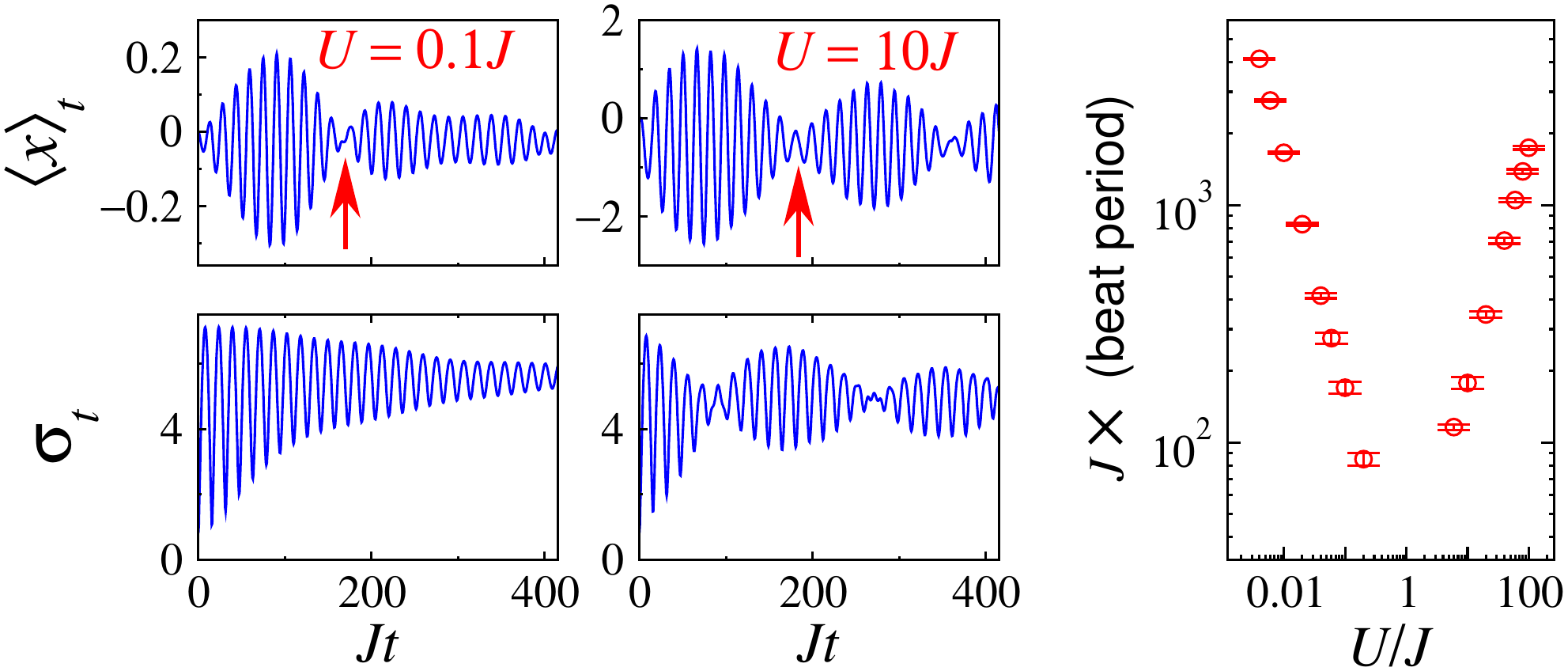}\tabularnewline
\end{tabular}
\par\end{centering}
\caption{\label{fig:beat_period} Beating/relaxation of Bloch oscillations near the free points. BHM,
  3 particles in 49 sites, $E=0.2J$, product initial state (\ldots01110\ldots).  Arrows indicate
  beat periods.
}
\end{figure}

\emph{Dynamics near `free' points---} Fig.~\ref{fig:beat_period} shows some BHM time
evolutions at finite interaction values near the `free' points $U=0,\infty$.  The center of
mass $\av{ x}_t = (\sum_j n_j\left(t\right) j) / N_p$ and the width
$\sigma_{t}=\left[\sum_{j}\left(j -\av x_{t}\right)^{2} n_j\left(t\right) / N_p \right]^{1/2}$ of
the cloud both generically show a ``collapse and revival'' or beating behavior.  Other cloud
characteristics (skewness or kurtosis) show the same effect \cite{supmat}.  To what extent the
phenomenon is visible varies with the initial state and the quantity observed, but generically for
$U/J$ not too close to $1$, a beat is visible.  The beat period is seen to have clear linear
dependences, $\propto{U}^{-1}$ at small $U/J$ and $\propto{U}$ at large $U/J$, on the interaction.
The behavior at small $U/J$ has recently been observed experimentally \cite{Meinert_PRL2014}.  We
have found the same behavior in the fermionic case as a function of $V$ \cite{supmat}.

This remarkably simple dependence can be explained using the many-body spectrum.  At the free
points, this spectrum is exactly equally spaced (steps of $E$) and highly degenerate.  As one moves
away from these simple points, the degeneracy is lifted, so that the frequencies available for the
dynamics are a range of values around $E$, the range being small compared to $E$.  This explains the
beat behavior.  A perturbative argument yields an energy level splitting of the order of $V^\nu$ or
$U^\nu$ with $\nu=\pm1$ for weak/strong interactions.  The splitting scale provides the beat
frequency.

Spectral considerations also explain why there is rapid relaxation behavior without beats in the $U,V \sim J$ regime.  In this regime, the eigenstates mix, destroying the ladder structure, and the chaotic structure of the spectrum leads to relaxation, as we have analyzed above.  The present study in terms of the spectrum thus explains the results of the experiments of Ref.~\cite{Meinert_PRL2014}.

\begin{figure*}[!t]
\includegraphics[width=2\columnwidth]{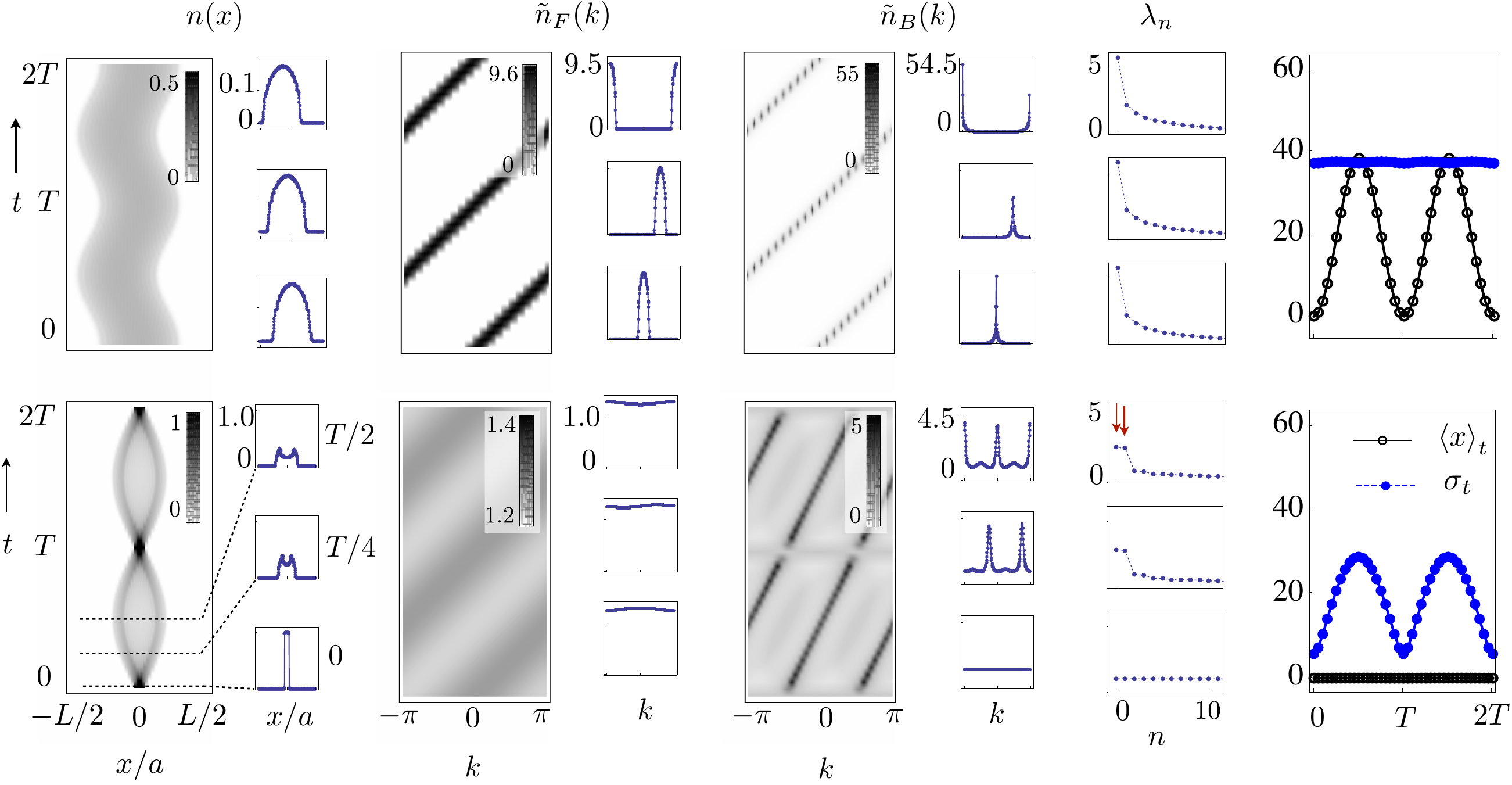}
\caption{\label{fig:small_rho}Time evolution of
  $n\left(x\right),\tilde{n}_{F}\left(k\right)=\sum_{x}\frac{1}{\sqrt{L}}\sum_{y}\rho_{y+x,y}^{F}e^{-ikx}$,
  $\tilde{n}_{B}\left(k\right)=\sum_{x}\frac{1}{\sqrt{L}}\sum_{y}\rho_{y+x,y}^{B}e^{-ikx}$ and
  $\lambda_{n}$ computed for $L=250;\, N_{p}=20;\, E=0.05J$.  Upper (lower) row: $\tilde{\rho}=0.1$
  ($\tilde{\rho}=10$).  The insets display the profiles of the different quantities
  for $t=T$, $t=T/4$ and $t=T/2$ where $T=2\pi/E$ is the oscillation period.  Right panels show
  evolution of the center of mass and width of the cloud. }
\end{figure*}

\emph{Cloud dynamics at `free' points---} In contrast with the equilibration seen for moderate
interactions, at $U,V=0,\infty$ there are perfectly periodic oscillations.  The long term state
is not equilibrated and has strong dependence on the initial condition.

Fig. \ref{fig:small_rho} shows time evolution for the JW-related cases $V=0$ and $U\to\infty$,
respectively labeled by F or B.  The cases of an initially spread-out and narrow cloud (small and
large $\tilde{\rho}$) are shown (top and bottom).  The density plots show the evolution of the
density $n_j(t)$ (identical for F and B), and of the momentum occupation number
$\tilde{n}_{F}\left(k,t\right)$ and $\tilde{n}_{B}\left(k,t\right)$.  For the bosonic system we also
compute the occupation numbers of the natural orbitals $\lambda_{n}(t)$
\cite{Penrose1956,Leggett2001} (with $\lambda_{0}\ge\lambda_{1}\ge...$), defined as the eigenvalues
of the single particle density matrix ${\rho_{\text B}}_{i,j}(t) = \av{ b^\dagger_i(t) b_j(t)} $.  A
macroscopic occupation (i.e., a $\lambda_{i}$ of order $\sqrt{N_{p}}$) corresponds to
quasi-condensation.
%
%  in the $\ket{\Psi_{i}}$ orbital, with $\rho_{\text B} \ket{\Psi_{i}} = \lambda_i \ket{\Psi_{i}}$.
%
%In one dimension no strict condensation arises for an Hamiltonian with short range interactions, instead there is the possibility of having quasi-condensed ground states with $\lambda_{0}\propto\frac{1}{\sqrt{N_{p}}}$ \cite{Rigol2004}.
% The insets display the profile of each quantity at $t=0,1/4T$ and $T/2$. 
% The right panels display the time evolution of the center of mass and of the cloud width. 

The density profile $n(x,t)$ displays qualitatively different dynamics for small and large
$\tilde{\rho}$: Bloch oscillations consist of mainly position oscillations for $\tilde{\rho}\ll1$
and mainly width oscillations for $\tilde{\rho}\gg1$.  For large $\tilde{\rho}$ the shape of the
initially localized cloud changes considerably within a period, the shape becoming double-peaked
when the cloud widens. The oscillation amplitude of the center of mass $\av x_{t}$ is large for
$\tilde{\rho}\ll1$ and small for $\tilde{\rho}\gg1$.  The cloud width $\sigma_{t}$ shows the
opposite behavior.  (Fig. \ref{fig:small_rho} right.)  This distinction is analogous to that
observed in single-particle Bloch oscillations \cite{Hartmann2004}.  Additional shape dynamics
appear at intermediate $\tilde{\rho}$ --- the cloud becomes strongly skewed once every period
\cite{supmat}.  The amplitude of skewness oscillations is non-monotonic as a function of
$\tilde{\rho}$, unlike amplitudes of position (width) oscillations which decreases (increases)
monotonically with $\tilde{\rho}$ \cite{supmat}.

%and the skewness $s_{t}=\frac{\sigma_{t}^{-3/2}}{N_{p}}\sum_{n}\left(x_{n}-\av x_{t}\right)^{3}n\left(x_{n},t\right)$
%that characterizes the asymmetry of the distribution $n\left(x,t\right)$.
%

For any Gaussian initial state, the subsequent cloud dynamics (time evolution of moments) can be obtained analytically as a function of the initial moments of correlators \cite{supmat}. 
%
% The knowledge of the exact solution allows to a systems of an arbitrary number of particles. 
%
The center of mass has purely sinusoidal oscillations, $\av
x_{t}=-\frac{2J}{E}\sin^{2}\left(\frac{tE}{2}\right)\mu_{1}$.  The width dynamics is more
complicated:
$\sigma_{t}^{2}-\sigma_{t=0}^{2} = -\frac{4J^{2}}{E^{2}}\sin^{4}\left(\frac{tE}{2}\right)\mu_{1}^{2}+2\left(\frac{J}{E}\right)^{2}\sin^{2}\left(\frac{E}{2}t\right)\left[1-\cos\left(Et\right)\mu_{2}\right]$.
Here $\mu_{a}=\frac{1}{2N_{p}}\sum_{y}\av{c_{y}^{\dagger}c_{y-a}+c_{y-a}^{\dagger}c_{y}}_{t=0}$.
(The behavior of $\mu_1$ and $\mu_2$ as functions of $N_p$ and $\tilde\rho$ is described in
\cite{supmat}.)  This allows to compute the amplitudes of oscillation of the moments, e.g, $\Delta
x=\max_{t}\av x_{t}-\min_{t}\av x_{t}$, and
$\Delta\sigma^{2}=\max_{t}\sigma_{t}^{2}-\min_{t}\sigma_{t}^{2}$ as a function of $\tilde{\rho}$ for
different $N_p$.  The position oscillation amplitude is $\Delta x=2J/E$ for $\tilde{\rho}\to0$, and
at large $\tilde{\rho}$ decreases as $\Delta x\propto J/\left(E\tilde{\rho}\right)$ for $N_{p}>1$
\cite{supmat}.
Conversely, $\Delta\sigma^{2}$ increases from zero to $2(J/E)^2$ as $\tilde{\rho}$  is increased ($N_{p}>1$) \cite{supmat}.    
%
% The dependence of $\Delta\sigma^{2}$ on $N_{p}$ and $\tilde{\rho}$ can be extracted from $\av x_{t}$
% and $\sigma_{t}^{2}$, and is given by $\Delta\sigma^{2}E^{2}/J^{2}=\max\left\{\frac{1}{4}\frac{\left(1-\mu_{2}\right){}^{2}}{\abs{\mu_{1}^{2}-\mu_{2}}}\right.$
% $,2\abs{\mu_{2}-2\mu_{1}^{2}+1},\left.\frac{1}{4}\frac{\left(3\mu_{2}-4\mu_{1}^{2}+1\right){}^{2}}{\abs{\mu_{1}^{2}-\mu_{2}}}\right\} $
% if $\abs{\frac{3\mu_{2}-4\mu_{1}^{2}+1}{2\left(\mu_{1}^{2}-\mu_{2}\right)}}<1$ and $\Delta\sigma^{2}E^{2}/J^{2}=2\abs{\mu_{2}-2\mu_{1}^{2}+1}$ otherwise.

The momentum distribution of the fermionic system (F) has simple time evolution:
$n_{F}(k,t)=n_{F}(k-\frac{2\pi}{T}t,0)$, reminiscent of single-particle Bloch oscillations
\cite{supmat}.
%
% \comm{(Why citing Supp?)}   
%
For B, the momentum distribution $n_{B}(k,t)$ has similar behavior for small $\tilde{\rho}$, but it
is now a sharply peaked distribution that traverses the Brillouin zone periodically, signaling
quasi-condensation in the initial state that survives during the oscillations.  The natural orbital
occupancy accordingly shows a dominant eigenvalue that stays dominant throughout the evolution.  The
large $\tilde{\rho}$ behavior is more intricate.  Although the condensate is initially
non-condensed, two well-defined coherence peaks appear.  Remarkably, they disappear periodically for
a short fraction of the period when returning to the initial state.  The set $\{\lambda_i\}$ now has
\emph{two} dominant occupancies, $\lambda_{0}\gg\lambda_{n>0}$, signaling a fragmented condensate
that is dynamically generated \cite{Rigol2004, Cai2011} and persists for almost all times
within each period.

% \textbf{Exact solution for the time evolution}

%the Bose-Hubbard Hamiltonian, $E$ the tilt strength and where the
%operator $b_{j}^{\dagger}$ creates of a boson on site $j$. The initial
%state, at $t=0$, is taken to be the ground-state of an harmonic trapped
%system described by the Hamiltonian $H_{0}=H_{\text{BH}}+W\sum_{j}j^{2}\, b_{j}^{\dagger}b_{j}.$

\emph{Discussion---}
We have presented a thorough study of many-body Bloch dynamics in two standard lattice models in one
dimension, one fermionic and one bosonic.
A main result is that generic many-body systems under a tilt potential have a dichotomic nature,
possessing both ETH and MBL features.  Although their eigenstates are exponentially localized, and
an initially trapped cloud has finite overlap only with a zero-measure set of eigenstates within the
relevant energy window, the long time dynamics yield a thermalized state within a Hilbert space of
effective dimension $d_{\text{eff}}$ which increases with the number of particles $N_p$.

The approach to the thermalized state can be seen as the destruction of the many-body Bloch
oscillations which are present at the integrable (`free') limits, both for weak and strong
coupling. We show that the relevant time-scale grows as $U$ ($U^{-1}$) or $V$ ($V^{-1}$) away from
the weak (strong) integrable limit.
At the free limits we present several striking features of the cloud dynamics, including a dynamical
generation (and periodic disappearance) of fragmented condensation for strong initial trapping.

%% For the zero or infinite interacting limits we explore explicit mappings to the non-interacting case
%% which allows to compute the cloud dynamics from the moments of the initial state distribution
%% function. This has allowed us to perform a detailed study of the cloud dynamics and, in particular
%% to identify a collective oscillation or a sequence of expansions and collapsed of the atomic cloud
%% depending on the initial state. Interestingly, we also found a dynamical generation (and periodic
%% disappearance) of fragmented condensation for box-like initial conditions (strong initial trapping).

%- differences between weak and strong initial trapping seen through first and second moments;

% - a rich set of relatively simple scaling behaviors;

%- dynamics of first and second moments expressed in terms of moments of the initial state;

% - asymmetric distortion effect + non-monotonic dependence of third moment on initial trapping strength;

% - dynamical generation (and periodic disappearance) of fragmented condensation for box-like initial conditions (strong initial trapping). 

% - $\Delta{x}$ and $\Delta{\sigma}$ at finite interactions --- how they change with trap strength.
%The basic physics of  $\Delta{x}$ increasing and  $\Delta{\sigma}$ decreasing with shallower traps, still survives.  Plots for t-V model.  Maybe include in Supplementary. 

\begin{acknowledgments}
P. Ribeiro acknowledges support by FCT through the Investigador FCT contract IF/00347/2014 and Grant No. UID/CTM/04540/2013.
\end{acknowledgments}

%\pagebreak{}

%\bibliographystyle{apsrev4-1}

%merlin.mbs apsrev4-1.bst 2010-07-25 4.21a (PWD, AO, DPC) hacked
%Control: key (0)
%Control: author (8) initials jnrlst
%Control: editor formatted (1) identically to author
%Control: production of article title (-1) disabled
%Control: page (0) single
%Control: year (1) truncated
%Control: production of eprint (0) enabled
%

\newpage

\begin{center}
  {\large
    Supplemental Materials for: \\
 \emph{Many-body quantum dynamics of initially trapped systems due to a Stark potential ---
  thermalization vs.\ Bloch oscillations}
}
\end{center}

%% \author{P.~Ribeiro}
%% \author{A.~Lazarides}
%% \author{Masudul Haque}

%% \affiliation{Max-Planck-Institut f\"ur Physik komplexer Systeme, N\"othnitzer Stra\ss e 38, 01187 Dresden, Germany}

%% \begin{abstract} 
%% In these Supplements, we 
%% %
%% \\ (1) do this and that
%% %
%% \\ (2) and also that and this
%% \end{abstract}

%%%%%%%%%%%%%%%%%%%%%%%%%%%%%%%%%%%%%%%%%%%%%%%%%%%%%%%
% Counter resetting for supplementaries
\setcounter{page}{1} \renewcommand{\thepage}{S\arabic{page}}

\setcounter{figure}{0}   \renewcommand{\thefigure}{S\arabic{figure}}

\setcounter{equation}{0} \renewcommand{\theequation}{S.\arabic{equation}}

\setcounter{table}{0} \renewcommand{\thetable}{S.\arabic{table}}

\setcounter{section}{0} \renewcommand{\thesection}{S.\Roman{section}}

\renewcommand{\thesubsection}{S.\Roman{section}.\Alph{subsection}}

% referring to subsections: don't prefix by section because that's already contained in
% \thesubsection.  Prefixes can be set using \p@*****.  Needs to be set within \makeatletter and
% \makeatother so that @ can be used in this way --->  

\makeatletter
\renewcommand*{\p@subsection}{}  
\makeatother

\renewcommand{\thesubsubsection}{S.\Roman{section}.\Alph{subsection}-\arabic{subsubsection}}

\makeatletter
\renewcommand*{\p@subsubsection}{}  % referring to subsubsections
\makeatother

%%%%%%%%%%%%%%%%%%%%%%%%%%%%%%%%%%%%%%%%%%%%%%%%%%%%%%%

\section{Contents\label{sec:Supplementary-Information}}

In these Supplemental Materials, 

\begin{itemize}[leftmargin=*]

\item We describe the four non-interacting (`free') points in the two one-dimensional models we have
  considered (Section \ref{suppsec_freepoints}).  These are the $U=0, \infty$ points for the Bose-Hubbard model (BHM) and the
  $V=0,\infty$ points for the fermionic t-V model (Ft-VM).

\item We provide a number of exact results valid at two of the free points, with derivations  (Section
  \ref{suppsec_analytic}). 

\item We discuss the beating behavior and provide further numerical data for the beating dynamics of
  the interacting systems (Section \ref{suppsec_beating}). 

\item We discuss some aspects of the many-body spectrum, including level-spacing statistics (Section
  \ref{suppsec_spectrum}).  The level-spacing data supports our picture of dual thermalizing and
  locallizing behavior.  

\end{itemize}

\section{The four `free' points} \label{suppsec_freepoints}

For the bosonic $U=0$ and the fermionic $V=0$ cases, the Hamiltonian is quadratic, i.e., that of
free bosons and free fermions respectively.  This implies that the knowledge of all $n$-point
correlators of the initial state allows for an analytical solution of the subsequent evolution of
the $n$-point correlators.  If the initial state is Gaussian (so that all $n$-point correlators are
determined by 2-point correlators), then the state remains Gaussian under time evolution.  This is
particularly straightforward for the $V=0$ case where all physically motivated initial states we
used are Gaussian, i.e. can be seen as ground states of free particle models.  

The non-interacting bosonic model ($U=0$) does not admit an initial Gaussian state at zero
temperature and finite particle number.  Nonetheless, from the knowledge of the correlation matrix
(matrix of two-point correlators) in the initial state one can obtain all subsequent two-point
correlators.  If the two-point correlators of the bosonic $U=0$ and of the fermionic $V=0$ are the
same initially, then in the subsequent dynamics the two-point correlators of the two models continue
to be the same.
  
The dynamics at the bosonic $U=\infty$ point can be mapped to that of the fermionic $V=0$ point by a
Jordan-Wigner (JW) transformation that provides a mapping at the operator level.  In particular, the
density of the fermionic and bosonic systems are the same.  (As a side product, this allows us to
conclude that the dynamics of the two-point correlators at $U=0$ and $U\to\infty$ is also the same
if the initial state is the same.)  Off-diagonal elements in position basis for the bosonic
 $U=\infty$ point and the fermionic $V=0$ point are not so simply related, but the hard-core boson
correlators can be computed numerically from the $V=0$ free fermion dynamics.  We have presented
such results in the main text.
 
Finally, although we can prove that the spectral properties of the $V=\infty$ are that of free
particles (next section), no mapping was found at the operator level and thus the dynamics of
correlators or densities is not simply related to that of a non-interacting model.

\subsection{$V\to\infty$ to $V=0$ mapping for the $t$-$V$ model}

The fermionic $t$-$V$ model allows for a mapping between the $V\to\infty$ Hamiltonian (in the reduced
Hilbert space sector where the energy is finite) and the $V=0$ Hamiltonian on a smaller chain.

In the $V\to\infty$ limit, states with particles at nearest-neighbour sites are kinematically
excluded.  For $N_p$ fermions in $L$ sites, this means an effective Hilbert space of dimension
$\begin{pmatrix}L-N_p+1\\ N_p\end{pmatrix}$, which is the same as the Hilbert space dimension of a
  system with $N_p$ fermions in $L-N_p+1$ sites without this constraint.

%%%%%%%%%%%%% FIGURE IN SUPPLEMENTARY %%%%%%%%%%%%%%%%%%%%%%%%%%%%%%%
\begin{figure}
\begin{centering}
\begin{tabular}{c}
\begin{tabular}{c}
\includegraphics[width=1\columnwidth]{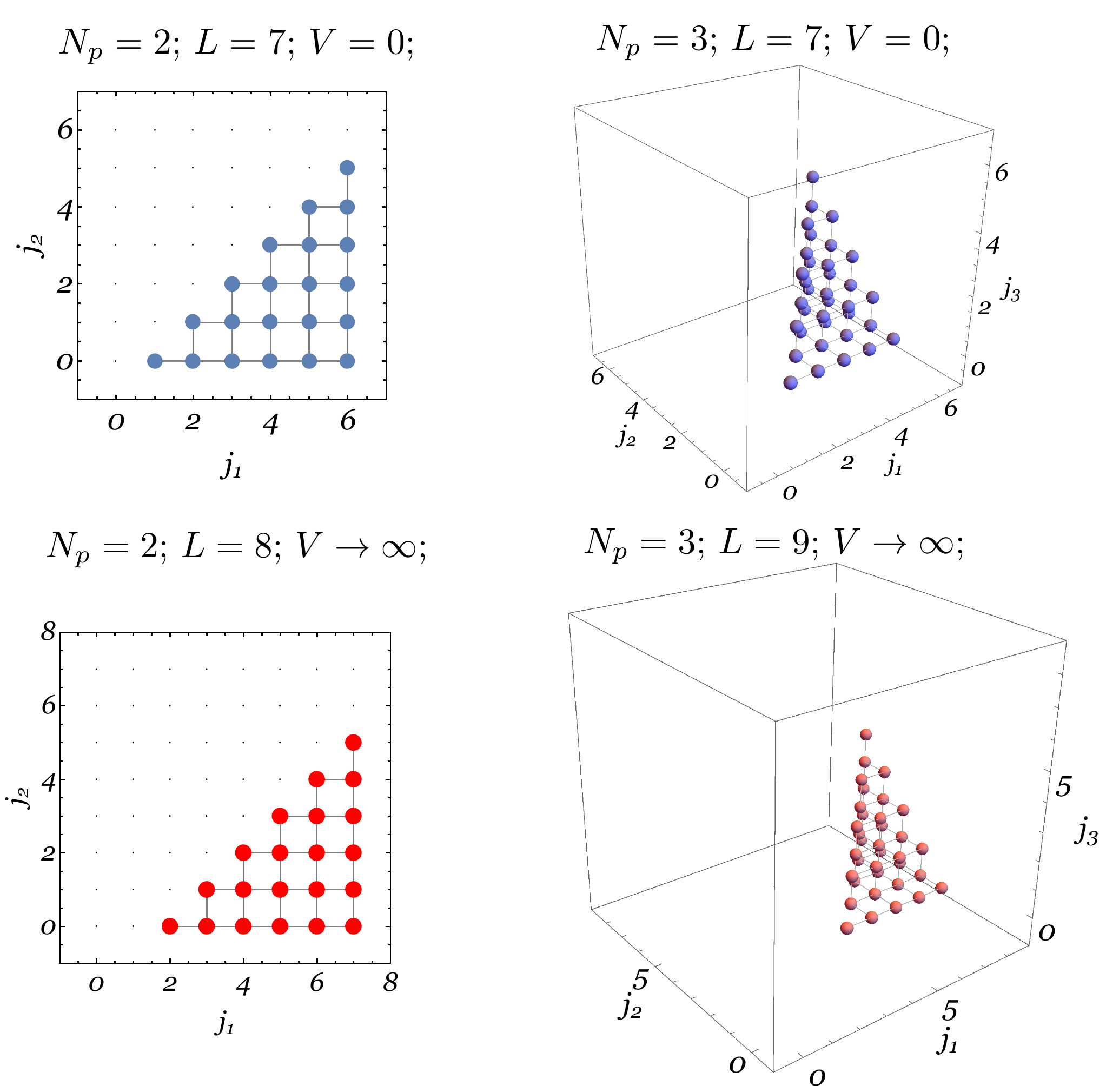}\tabularnewline
\end{tabular}\tabularnewline
\end{tabular}
\par\end{centering}
\caption{\label{fig:V0vsVinf} Mapping between Hilbert spaces of the $V=0$ fermionic system and the
  $V\to\infty$ fermionic system with $N_p-1$ more sites on the tight-binding chain.  Left panels:
  $N_p=2$ particles.  The allowed configurations of the $V=0$ system are shown on the $(j_1,j_2)$
  plane (left top), where $j_1$ and $j_2$ are the positions of the two particles.  Lines indicate
  configurations connected by a single-particle hopping process.  Left bottom panel shows that the
  $V\to\infty$ system with one more lattice site has configuration space of the same size and same
  hopping topology.  Right panels: $N_p=3$ particles.  The corresponding $V\to\infty$ system now has
  two extra lattice sites compared to the $V=0$ system.
}
\end{figure}
%%%%%%%%%%%%% FIGURE IN SUPPLEMENTARY %%%%%%%%%%%%%%%%%%%%%%%%%%%%%%%

We can design a mapping of the Hilbert space for $V\to\infty$ for a chain of size $L$ and $N_p$
fermions onto the Hilbert space for $V=0$ with $N_p$ fermions on $L'=L-N_p+1$ sites.  The mapping
preserves the Hamiltonian.  The mapping of the Hilbert spaces is shown in Figure \ref{fig:V0vsVinf}
for $N_p=2$ and $3$, by showing the configuration spaces allowed in the two cases.  

For a 1D system with $N_p$ fermionic particles, anti-symmetry dictates that the configuration space
consists of states with $j_1<j_2<...<j_{N_p}$.  Thus, for $N_p=2$ particles where configurations can
be described as $(j_1,j_2)$ pairs, the allowed configurations exclude the diagonal line on the
$j_1$-$j_2$ plane.  In the $V\to\infty$ case, the states with particles at nearest-neighbour sites
$j_i$ and $j_i+1$ are also excluded; hence the next-to-diagonal points on the $j_1$-$j_2$ plane are
also excluded.  This results in an identical number and topology of allowed configurations in the
$V=0$ case and $V\to\infty$ cases, when there is one more site in the latter case.  This is
displayed in the left panels of Figure \ref{fig:V0vsVinf}.

The description is analogous for $N_p=3$, with the diagonal plane being excluded due to
antisymmetry, and one further next-to-diagonal plane being excluded in the case of $V\to\infty$, as
shown in the right panels of Figure \ref{fig:V0vsVinf}.  The construction is trivially generalized
to arbitrary $N_p< L/2$, but difficult to display visually for larger $N_p$.   In Table
\ref{fig:V0vsVinf_tabular_Np4}, the mapping is shown to work for a $N_p=4$ case, by listing all
the configurations in the two cases.   

\begin{table}
\begin{tabular}{c|c}
 $N_p=4$, $L=6$, $V=0$  &  $N_p=4$, $L=9$, $V\to\infty$  
\\
\hline
\\
0 0 1 1 1 1 &  0 0 1 0 1 0 1 0 1
\\
0 1 0 1 1 1 &  0 1 0 0 1 0 1 0 1
\\
0 1 1 0 1 1 &  0 1 0 1 0 0 1 0 1
\\
0 1 1 1 0 1 &  0 1 0 1 0 1 0 0 1
\\
0 1 1 1 1 0 &  0 1 0 1 0 1 0 1 0
\\
1 0 0 1 1 1 &  1 0 0 0 1 0 1 0 1
\\
1 0 1 0 1 1 &  1 0 0 1 0 0 1 0 1
\\
1 0 1 1 0 1 &  1 0 0 1 0 1 0 0 1
\\
1 0 1 1 1 0 &  1 0 0 1 0 1 0 1 0
\\
1 1 0 0 1 1 &  1 0 1 0 0 0 1 0 1
\\
1 1 0 1 0 1 &  1 0 1 0 0 1 0 0 1
\\
1 1 0 1 1 0 &  1 0 1 0 0 1 0 1 0
\\
1 1 1 0 0 1 &  1 0 1 0 1 0 0 0 1
\\
1 1 1 0 1 0 &  1 0 1 0 1 0 0 1 0
\\
1 1 1 1 0 0 &  1 0 1 0 1 0 1 0 0
\end{tabular}
\caption{\label{fig:V0vsVinf_tabular_Np4} 
Same mapping as in Figure \ref{fig:V0vsVinf}, now for $N_p=4$ particles.  The configuration space
would be 4-dimensional in the representation used in Figure \ref{fig:V0vsVinf}, so we simply list
all allowed configurations in the two cases.  It is easy to verify that: (1) if a pair of
configurations on the left are connected by single-particle hopping, then the corresponding pair on
the right is also connected by single-particle hopping; (2) once a zero energy is chosen for the
electric field for the two systems, the energy difference between a configuration on the left and
the corresponding configuration on the right is the same for all 15 configurations.
}
\end{table}

The matrix elements of the Hamiltonian between basis states correspond to the hopping of a single
particle.  Diagonal matrix elements are given by the potential energy of the tilting field.  The
Hamiltonians are identical in the case of the $V\to\infty$ system and the $V=0$ system with the
number of sites reduced by $N_p-1$, except for a possible constant shift on the diagonal terms.
(Since the lattice sizes are unequal in the two cases, the definition of zero Stark energy might be
chosen independently in the two cases, so this shift is arbitrary, and anyway does not affect the
dynamics.)  The off-diagonal matrix elements due to single-particle hopping are identical because
the mapping preserves the topology of the configuration spaces, i.e., if two configurations of the
$L$-site $V\to\infty$ system are connected by a single-particle hopping process, then the
corresponding configurations of the $(L-N_p+1)$-site $V=0$ system are also connected by a
single-particle hopping process.  In Figure \ref{fig:V0vsVinf} for the $N_p=2$ case, such pairs are
joined by lines, and it is visually obvious that the network topology is preserved under the mapping
of Hilbert spaces.

The present mapping provides a one-to-one correspondence between basis states and establishes the
equality of the Hamiltonians in the two cases.  However it does not translate to a simple mapping
between creation and annihilation operators, which would have allowed a computation of correlators
in the $V=\infty$ case from a free-particle calculation.  We are not aware of a mapping at the
operator level that takes us from the $V=\infty$ model to a non-interacting system.

\section{Exact solution at non-interacting points \label{suppsec_analytic}} 

In the main text, we highlighted some results for the fermionic $V=0$ and bosonic $U\to\infty$
systems.  We now provide some more details and explicit expressions for time evolution and
asymptotic behaviors.

\subsection{Analytic expressions for moments of the cloud}

We present the derivations for free fermions.  Since this concerns occupancies, the $U\to\infty$
bosonic system is described by the same equations.

The Hamiltonian can be  written as $H=\mathcal T+ \mathcal
E$ with
\begin{eqnarray}
\mathcal T & = & \int_{-\pi}^{\pi}\frac{dk}{2\pi}c_{k}^{\dagger}\varepsilon\left(k\right)c_{k}\\
\mathcal E & = & \sum_{n=-\infty}^{\infty}En\, c_{n}^{\dagger}c_{n}
\end{eqnarray}
where the operators in real and momentum space obey the usual relations
$c_{n}=\int_{-\pi}^{\pi}\frac{dk}{2\pi}e^{ikn}c_{k}$, $c_{k}=\sum_{n}e^{-ikn}c_{n}$. For
nearest-neighbor hoppings, $\varepsilon\left(k\right)=-J\cos\left(k\right)$, however the following
argument holds for a generic dispersion relation. Using the fermionic commutation relations, the
evolution operator can be written as
\begin{eqnarray}
e^{-iHt} & = &
e^{-i\int_{-\pi}^{\pi}\frac{dk}{2\pi}c_{k}^{\dagger}\left[\int_{0}^{t}dt'\,\varepsilon\left(k+Et'\right)\right]c_{k}}e^{-i 
  \mathcal E t} .
\end{eqnarray}
Applying the evolution operator in this form to the single-particle density matrix in momentum space
yields 
\begin{multline}
\av{c_{k}^{\dagger}c_{k'}}_{t} ~=~
e^{-i\int_{0}^{t}dt'\,\left[\varepsilon\left(k'+Et'\right)-\varepsilon\left(k+Et'\right)\right]}
\\
\times \av{c_{\left(k+E t\right)}^{\dagger}c_{\left(k'+E t\right)}}_{t=0}\label{eq:rho_kkp}
\end{multline}
where $\av{...}_{t}$ denotes the mean value taken at time $t$. For
$\varepsilon\left(k\right)=-J\cos\left(k\right)$, Eq.(\ref{eq:rho_kkp})
simplifies to 
\begin{multline}
\av{c_{k}^{\dagger}c_{k'}}_{t} ~=~
e^{2i\frac{J}{E}\sin\left(\frac{Et}{2}\right)\left[\cos\left(k'-\frac{tE}{2}\right)-\cos\left(k-\frac{Et}{2}\right)\right]}
\\ \times \av{c_{\left(k+Et\right)}^{\dagger}c_{\left(k'+Et\right)}}_{t=0}\label{eq:rho_kkp-1}
\end{multline}
so that the real-space correlators are found to be
\begin{multline} \label{eq:sol}
\av{c_{x}^{\dagger}c_{x'}}_{t}=\sum_{y,y'}\left\{ e^{i\frac{Et}{2}\left(x+y\right)}I_{x-y}\left[-2i\frac{J}{E}\sin\left(\frac{Et}{2}\right)\right]\right\} \times\\
\av{c_{y}^{\dagger}c_{y'}}_{0}\left\{
e^{-i\frac{Et}{2}\left(x'+y'\right)}I_{-x'+y'}\left[2i\frac{J}{E}\sin\left(\frac{Et}{2}\right)\right]\right\}  
\end{multline}
using the identity $e^{z\cos\left(\theta\right)}=\sum_{n=-\infty}^{\infty}I_{n}\left(z\right)e^{in\theta}$
where $I_{n}\left(z\right)$ is the modified Bessel function. 

In order to compute the moments of the cloud we define the generalized
characteristic function 
\begin{eqnarray}
G_{a}\left(\lambda,t\right) & = & \sum_{y}e^{i\lambda\left(y-\frac{a}{2}\right)}\av{c_{y}^{\dagger}c_{y-a}}_{t}\nonumber \\
 & = & \sum_{m}\frac{\left(i\lambda\right)^{m}}{m!}N_{p}^{m+1}\mu_{a,m}\left(t\right)\label{eq:G_a_0}
\end{eqnarray}
with 
\begin{eqnarray}
\mu_{a,m}\left(t\right) & = & N_{p}^{-\left(m+1\right)}\sum_{y}\left(y-\frac{a}{2}\right)^{m}\av{c_{y}^{\dagger}c_{y-a}}_{t}
\end{eqnarray}
the generalized moments.

\begin{figure}
\begin{centering}
\includegraphics[width=1\columnwidth]{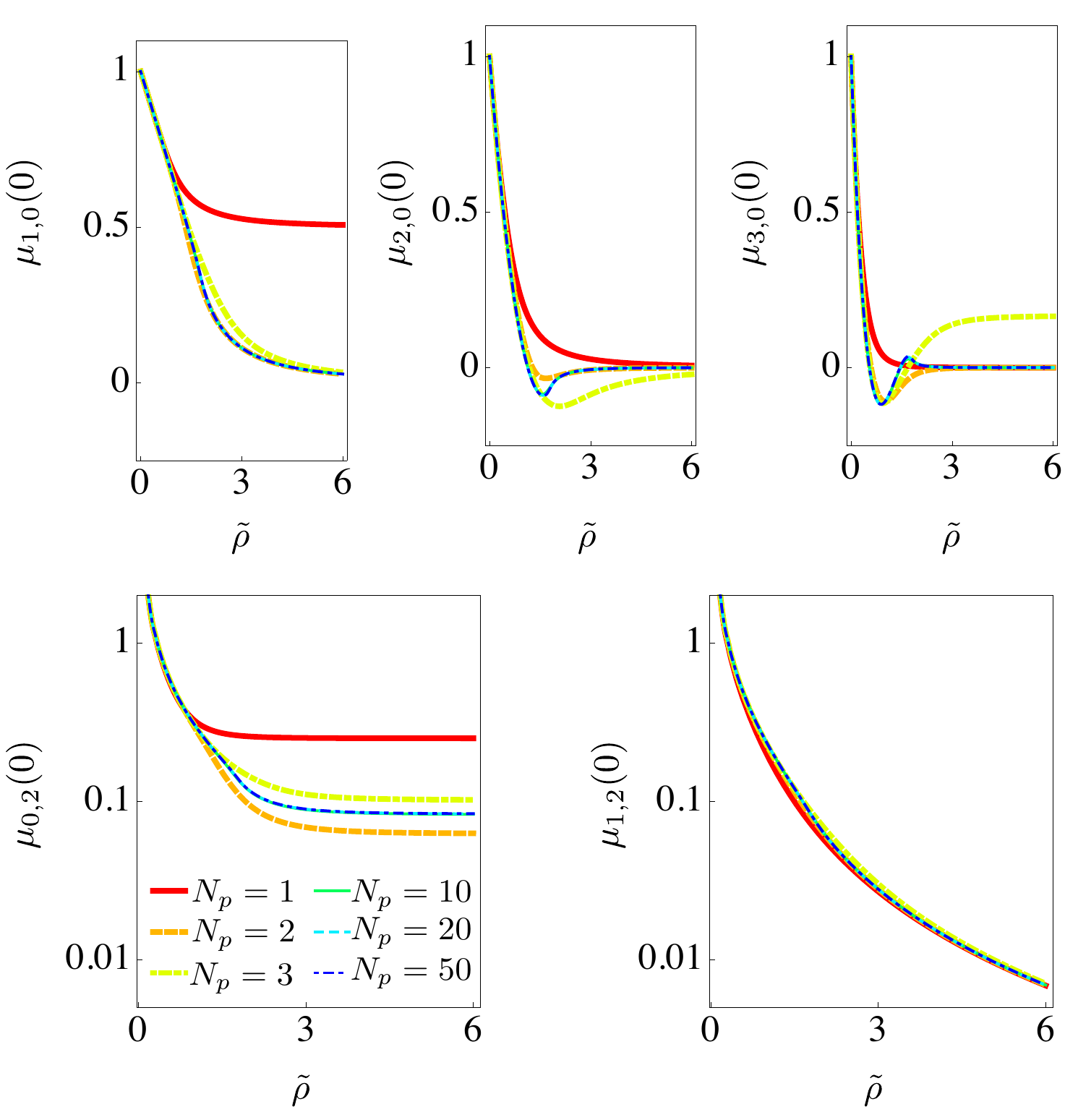}\tabularnewline
\end{centering}
\caption{\label{fig:mu_s} Numerical evaluation of the generalized moments
$\mu_{a,m}$ of the ground-state of the harmonic trap as a function
of $\tilde{\rho}$ for several values of $N_{p}$. }
\end{figure}

For $a=0$, the $\mu_{a,m}$ are simply the moments of the cloud shape:
$\av{x^{m}}_{t}=N_{p}^{m}\mu_{0,m}\left(t\right)$.  For $a\neq0$, they may be regarded as moments of
two-point correlators.  In particular, the quantities $\mu_{1}$ and $\mu_{2}$ defined in the main
text equal to $\mu_{1,0}\left(t=0\right)$ and $\mu_{2,0}\left(t=0\right)$ respectively. In case
the initial state is the ground-state of an harmonic trap one has
$\mu_{-a,m}\left(0\right)=\mu_{a,m}\left(0\right)\in\mathbb{R}$ for $m$ even and
$\mu_{a,m}\left(0\right)=0$ for $m$ odd. The first non-trivial generalized moments
$\mu_{a,m}\left(t\right)$ of the trap ground state are shown in Fig.\ref{fig:mu_s} as function of
$\tilde{\rho}$, for different values of $N_{p}$.

Using Eq.(\ref{eq:sol}), the characteristic function can be written as
\begin{multline}
G_{a}\left(\lambda,t\right)=e^{iEta}\sum_{b}e^{-i\left(\frac{E}{2}t+\frac{\pi}{2}\right)b}\\
\times I_{b}\left[4i\frac{J}{E}\sin\left(\frac{Et}{2}\right)\sin\left(\frac{\lambda}{2}\right)\right]G_{a-b}\left(\lambda,0\right)\label{eq:G_a_1}
\end{multline}
Taylor expanding the previous expression and identifying the powers of $\lambda$ in both sides of
Eqs. (\ref{eq:G_a_0}) and (\ref{eq:G_a_1}) for the case $a=0$, one obtains explicit expressions for
the first 3 moments of the cloud shape:
\begin{equation*}
\mu_{0,1}\left(t\right)=N_{p}^{-1}\av
x=-\frac{2}{N_{p}}\frac{J}{E}\sin^{2}\left(\frac{Et}{2}\right)\mu_{1,0}\left(0\right) , 
\end{equation*}
\begin{multline*}
\mu_{0,2}\left(t\right)=N_{p}^{-2}\av{x^{2}}=\mu_{0,2}\left(0\right)\\
+\frac{2}{N_{p}^{2}}\left(\frac{J}{E}\right)^{2}\sin^{2}\left(\frac{Et}{2}\right)\left[1-\cos\left(Et\right)\mu_{2,0}\left(0\right)\right] ,
\end{multline*}
and
\begin{multline*}
\mu_{0,3}\left(t\right)=N_{p}^{-3}\av{x^{3}}=-\frac{6}{N_{p}}\frac{J}{E}\sin^{2}\left(\frac{Et}{2}\right)\mu_{1,2}\left(0\right)\\
-\frac{1}{N_{p}^{3}}\left(\frac{J}{E}\right)^{3}\sin^{2}\left(\frac{Et}{2}\right)\left\{ \left[6\sin^{2}\left(\frac{Et}{2}\right)+\frac{1}{2}\left(\frac{E}{J}\right)^{2}\right]\right.\\
\left.\times\mu_{1,0}\left(0\right)+\left[\cos\left(2Et\right)-\cos\left(Et\right)\right]\mu_{3,0}\left(0\right)\right\}  .
\end{multline*}
as a function of the initial values $\mu_{a,m}\left(0\right)$ of the generalized moments.  In the
cases of interest here (starting with trap ground states), these initial values are displayed in
Fig.\ref{fig:mu_s}.

\subsection{Cloud shape dynamics}

Using the exact solution above, one can describe the position and shape oscillations of the cloud
during the periodic evolution.  

\begin{figure}
\begin{centering}
\begin{tabular}{c}
\includegraphics[width=.7\columnwidth]{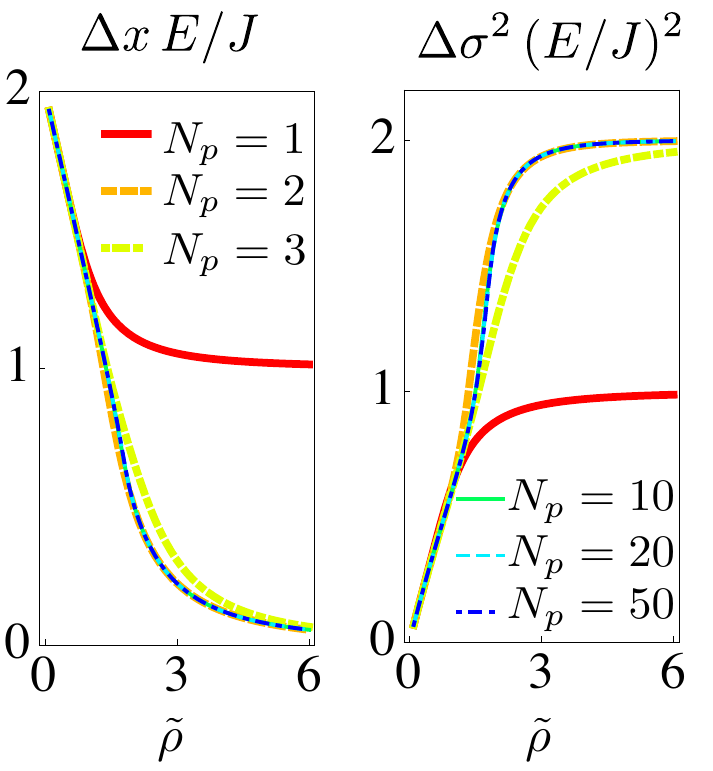}\tabularnewline
\end{tabular}
\par\end{centering}
\caption{\label{fig:Max_amp}Rescaled oscillation amplitudes of the center
of mass $\Delta x=\max_{t}\av x_{t}-\min_{t}\av x_{t}$ and cloud
width $\Delta\sigma^{2}=\max_{t}\sigma_{t}^{2}-\min_{t}\sigma_{t}^{2}$
as a function of $\tilde{\rho}$ for different values $N_{P}$. Note
that the rescaled quantities are independent of $E$.  }
\end{figure}

We consider the amplitude of variation of the center of mass
\[
\Delta x=\max_{t}\av
x_{t}-\min_{t}\av x_{t}
\]
and the amplitude of cloud width oscillations
\[
\Delta\sigma^{2}=\max_{t}\sigma_{t}^{2}-\min_{t}\sigma_{t}^{2}  . 
\]
Fig. \ref{fig:Max_amp} shows these two quantities as a function of $\tilde{\rho}$ for different
numbers of particles.  The quantities plotted, $\Delta{x}(E/J)$ and $\Delta\sigma^{2}(E/J)^2$, are
scaled to be unit-less and independent of the tilt $E$.  

The $\tilde{\rho}$-dependence (for both $\Delta{x}(E/J)$ and $\Delta\sigma^{2}(E/J)^2$) are very
similar for all $N_p>1$, converging rapidly to the large-$N_p$ limit.  The
single-particle ($N_p=1$) behavior differs significantly.

For $N_p>2$, $\Delta{x}(E/J)$ decreases from 2 to 0, while $\Delta\sigma^{2}(E/J)^2$ increases from
0 to 2.  This reflects the physics that the Bloch oscillations are primarily position oscillations
for small $\tilde{\rho}$ and primarily width oscillations for large $\tilde{\rho}$.  (For the
single-particle case, the large-$\tilde{\rho}$ limit is different.)

From the exact solutions, $\Delta{x}(E/J)$ and $\Delta\sigma^{2}$ can be expressed in terms of
$\mu_1=\mu_{1,0}(0)$ and $\mu_2=\mu_{2,0}(0)$.
It is easy to see that $\Delta{x}(E/J) = 2\mu_1$.
The expression for $\Delta\sigma^{2}$ is more complicated: $\Delta\sigma^{2}E^{2}/J^{2}=\max\left\{
\frac{1}{4}\frac{\left(1-\mu_{2}\right){}^{2}}{\abs{\mu_{1}^{2}-\mu_{2}}}, \xi,
\frac{\left(3\mu_{2}-4\mu_{1}^{2}+1\right){}^{2}}{4\abs{\mu_{1}^{2}-\mu_{2}}}\right\} $ if
$\abs{\frac{3\mu_{2}-4\mu_{1}^{2}+1}{2\left(\mu_{1}^{2}-\mu_{2}\right)}}<1$ and
$\Delta\sigma^{2}E^{2}/J^{2}=\xi$ otherwise. Here $\xi=2\abs{\mu_{2}-2\mu_{1}^{2}+1}$.

%%%%%%%%%%%%% FIGURE IN SUPPLEMENTARY %%%%%%%%%%%%%%%%%%%%%%%%%%%%%%%
%%%%%%%%%%%%% FIGURE IN SUPPLEMENTARY %%%%%%%%%%%%%%%%%%%%%%%%%%%%%%%

\begin{figure}
\begin{centering}
\begin{tabular}{c}
\includegraphics[width=1\columnwidth]{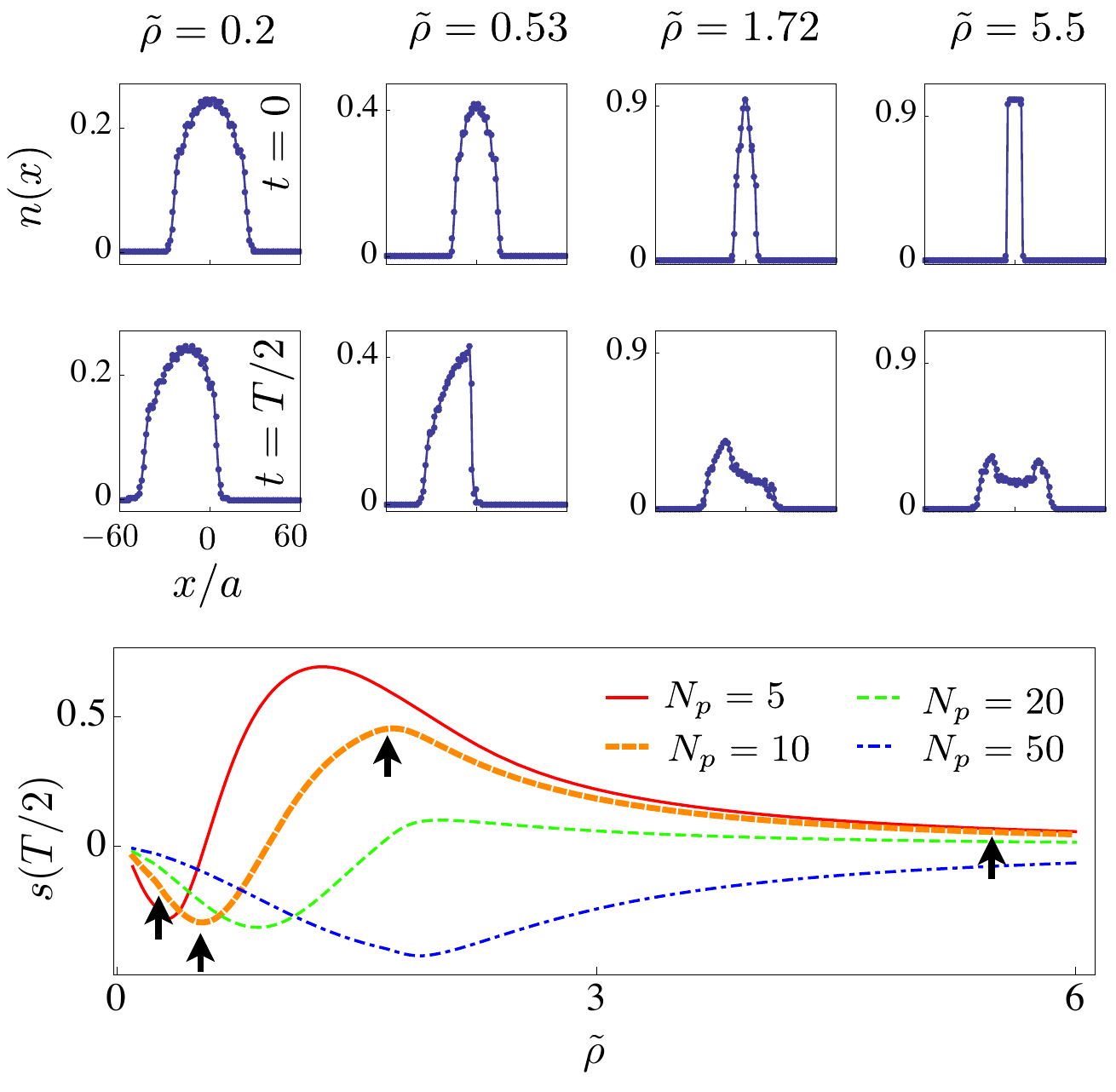}\tabularnewline
\end{tabular}
\par\end{centering}
\caption{\label{fig:Skew} Upper panels - Cloud shape at $t=0$ and $t=T/2$ for several values of
  $\tilde{\rho}$ and $N_{p}=10$.  Lower panel - Skewness of the atomic cloud at half a period
  plotted against $\tilde{\rho}$ for different values of $N_{b}$ computed for $E=0.1J$.  The
  $\tilde{\rho}$ values used in the upper panels are shown with arrows in the lower panel. 
}
\end{figure}

We now consider the third moment, which gives the skewness of the cloud.  
In Fig.(\ref{fig:Skew}) we show the skewness computed at $t=T/2$, where the cloud typically shows a
larger deformation with respect to its initial shape.  The skewness it seen to have a non-monotonic
dependence on $\tilde{\rho}$.  $s\left(T/2\right)$ vanishes for both $\tilde{\rho}=0$ and
$\tilde{\rho}\to\infty$.  For $\tilde{\rho}=0$ this is due to an almost undeformed cloud evolution.
In the $\tilde{\rho}\to\infty$ case, while there is significant deformation, the cloud remains
symmetric throughout the whole oscillation period.  For a fixed tilt strength $E$ and a large number
of particles (see Fig.(\ref{fig:Skew}) lower panel for $E=0.1J$ and $N_{p}=50$) this quantity passes
by an $N_{p}$-dependent minimum.

Fig. \ref{fig:Skew} upper panel shows the the density profile of the atomic cloud for $t=0$ and
$t=T/2$ for the points marked (with arrows) in the lower panel.  The minimum skewness point
corresponds to a highly asymmetric cloud shape having a shock-wave-like form (e.g., the
$\tilde{\rho}=0.53$ panel in Fig. \ref{fig:Skew}).  For smallish particle numbers
($N_{p}\lesssim10$), the $\tilde{\rho}$-dependence is more intricate --- there is both a positive
maximum and a negative minimum of $s(T/2)$.  The shape of the distorted cloud when having a positive
maximum is exemplified in the $\tilde{\rho}=1.72$ panel. 
For $N_{p}\gtrsim20$ there is a unique (negative) minimum that shifts to larger $\tilde{\rho}$ with
increasing $N_p$.

%%%%%%%%%%%%% FIGURE IN SUPPLEMENTARY %%%%%%%%%%%%%%%%%%%%%%%%%%%%%%%
%%%%%%%%%%%%% FIGURE IN SUPPLEMENTARY %%%%%%%%%%%%%%%%%%%%%%%%%%%%%%%

\subsection{Recurrent occupancies of the natural orbitals}

\begin{figure}
\begin{centering}
\begin{tabular}{c}
\includegraphics[width=1\columnwidth]{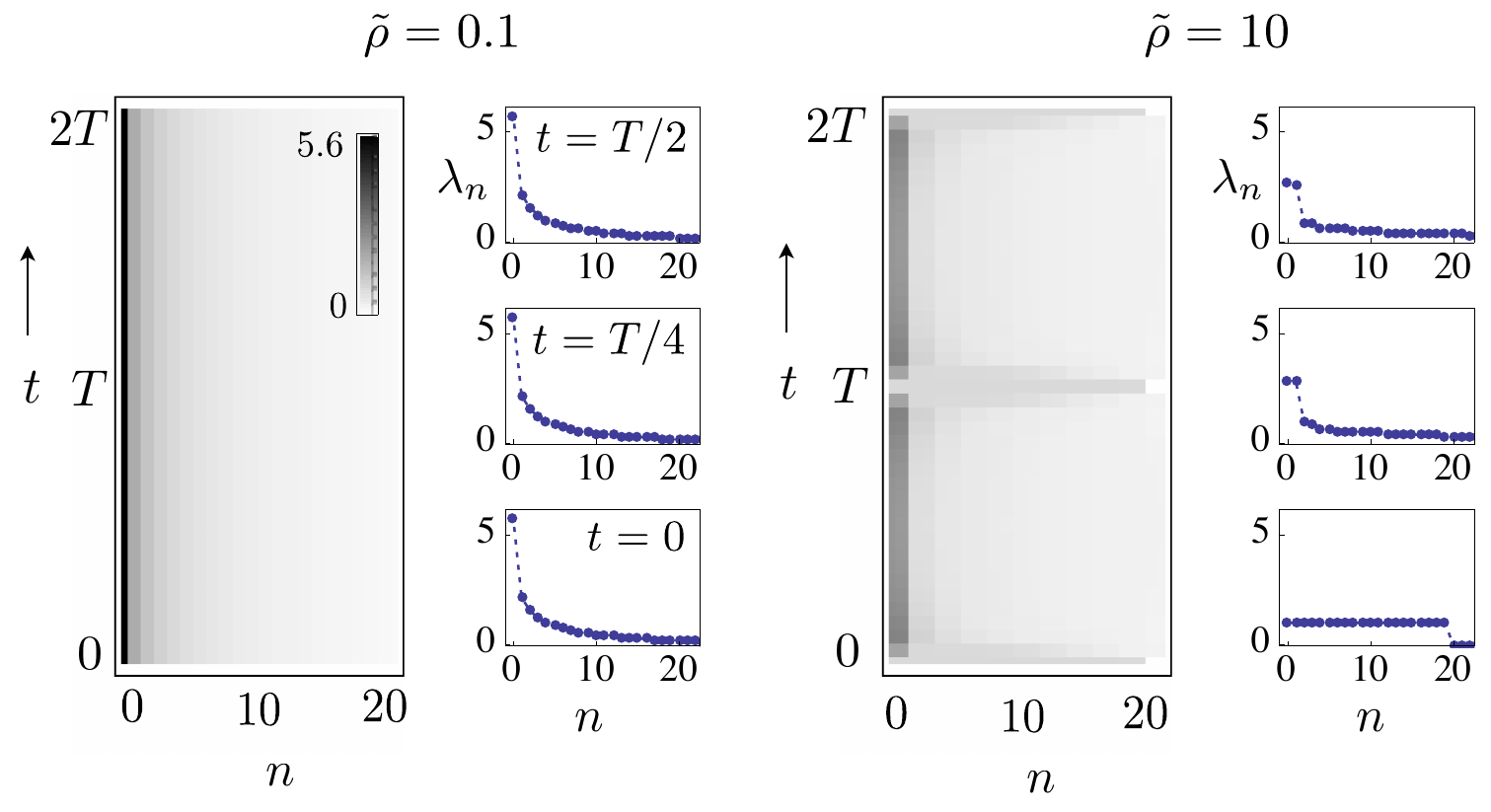}\tabularnewline
\end{tabular}
\par\end{centering}
\caption{\label{fig:Lambdas} Time evolution of the occupancies of natural orbitals for a system of
  $N_{p}=20$ HC bosons on a tilted lattice with $E=0.05J$, for the two different initial conditions
  discussed in the main text. }
\end{figure}

Here we present some more details of the time evolution of the natural orbital occupancies,
$\lambda_{n}$.  Figure \ref{fig:Lambdas} shows the $\lambda_{n}$ as a function of $n$, corresponding
to the two cases presented in Fig.3 in the main text, through density plots in addition ot
snapshots. 

For the $\tilde{\rho}=0.1$ case the lowest natural orbital has a substantial occupation already in
the initial state --- this is a single-mode quasi-condensate.  During the course of the evolution
the distribution of the $\lambda_{n}$'s does not get substantially modified and the initially
quasi-condensed state is observed to remain stable throughout the time evolution.

On the contrary, large $\tilde{\rho}$ induces a Mott insulator state as the initial condition for
which the occupation of the natural orbits is given by $\lambda_{n}=1$ for $n<N_{p}$ and
$\lambda_{n}=0$ otherwise.  In Figure \ref{fig:Lambdas} (right panels), we emphasize that this state
is realized periodically but only for times in a small vicinity of the multiples of the period
$t=mT$ with $m\in\mathbb{Z}$.  During most of the evolution the two lowest modes get substantially
occupied giving rise to a periodically regenerated bimodal quasi-condensate,  as described in the main text.

%%%%%%%%%%%%% FIGURE IN SUPPLEMENTARY %%%%%%%%%%%%%%%%%%%%%%%%%%%%%%%
%%%%%%%%%%%%% FIGURE IN SUPPLEMENTARY %%%%%%%%%%%%%%%%%%%%%%%%%%%%%%%

\subsection{Asymptotic behaviors --- density profile and fluctuations}

%%%%%%%%%%%%% FIGURE IN SUPPLEMENTARY %%%%%%%%%%%%%%%%%%%%%%%%%%%%%%%
\begin{figure}
\begin{centering}
\includegraphics[width=0.5\textwidth]{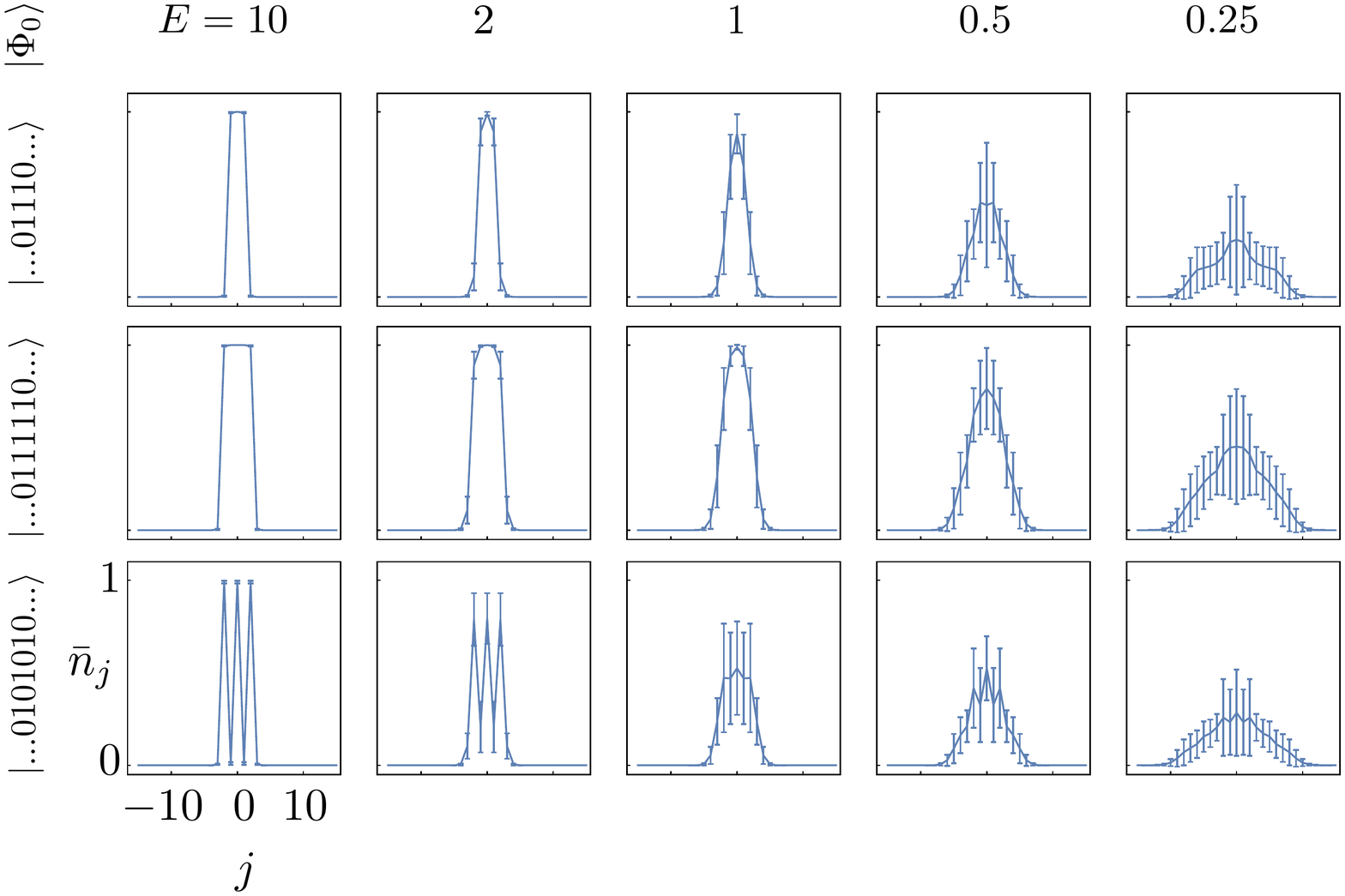}\tabularnewline
\end{centering}
\caption{\label{fig:n_j}Asymptotic time average values of the densities $\bar n_j$ computed from the
  exact solution, for three different initial states and various values of $E$.  We also show the
  fluctuations $\bar \sigma_j$, depicted as error bars.  }
\end{figure}
%%%%%%%%%%%%% FIGURE IN SUPPLEMENTARY %%%%%%%%%%%%%%%%%%%%%%%%%%%%%%%

The exact solution also allows to compute the long time asymptotic dynamics.  
From Eq.(\ref{eq:sol}),  the infinite time average value of the two-point functions is given by
\begin{multline}
  \overline{\av{c_{j}^{\dagger}c_{j'}}} ~=~
  \lim_{t\to\infty}\frac{1}{t}\int_{0}^{t}\av{c_{j}^{\dagger}c_{j'}}_{t'}dt'
  \\
  = \sum_{n,n'}M_{jj';nn'}\av{c_{n}^{\dagger}c_{n'}}_{0}.
\end{multline}
Since  the evolution is periodic, in the quantities $M_{jj';nn'}$ the integral can be taken over a period of the evolution 
\begin{multline}
M_{jj';nn'}=\frac{1}{T}\int_{0}^{T}dt\left\{ e^{i\frac{V_{0}t}{2}\left(j+n\right)}  \right.
\\ \left. 
\times   I_{j-n}\left[-2i\frac{J}{V_{0}}\sin\left(\frac{tV_{0}}{2}\right)\right] \right.
\\ \left. 
\times e^{-i\frac{V_{0}t}{2}\left(j'+n'\right)} \right.
\\ \left. 
\times
I_{-j'+n'}\left[2i\frac{J}{V_{0}}\sin\left(\frac{tV_{0}}{2}\right)\right] \right\} . 
\end{multline}
This derivation is completely general for an initial state correlation matrix $\av{c_{n}^{\dagger}c_{n'}}_{0}$. 

For simplicity, let us concentrate on the asymptotic form of the density, for which $j=j'$, for the
special case of initial condition with
$\av{c_{n}^{\dagger}c_{n'}}_{0}=\delta_{nn'}\av{c_{n}^{\dagger}c_{n}}_{0}$, i.e., product states as
initial states.  
For this particular case the long time averaged density yields
\begin{equation}
\bar{n}_{j}=\overline{\av{c_{j}^{\dagger}c_{j}}}=\sum_{a}m_{a}\av{c_{j+a}^{\dagger}c_{j+a}}_{0}
\end{equation}
with 
\begin{multline}
m_{a}=M_{jj;(j+a)(j+a)}\\=\frac{1}{T}\int_{0}^{T}dt\left(-1\right)^{a}\left\{ I_{a}\left[-2i\frac{J}{V_{0}}\sin\left(\frac{tV_{0}}{2}\right)\right]\right\} ^2
\end{multline}
In the same way, the fluctuation around the average are given by 
\begin{multline}
\bar{\sigma}_{j}^{2}=\overline{\left(\av{c_{j}^{\dagger}c_{j}}-\overline{\av{c_{j}^{\dagger}c_{j'}}}\right)^{2}}\\=\sum_{a,b}m_{a,b}\av{c_{j+a}^{\dagger}c_{j+a}}_{0}\av{c_{j+b}^{\dagger}c_{j+b}}_{0}-\bar{n}_{j}^{2}
\end{multline}
with 
\begin{multline}
m_{a,b}=\frac{1}{T}\int_{0}^{T}dt\left(-1\right)^{a+b} \times \\\left\{ I_{a}\left[-2i\frac{J}{V_{0}}\sin\left(\frac{tV_{0}}{2}\right)\right]I_{b}\left[-2i\frac{J}{V_{0}}\sin\left(\frac{tV_{0}}{2}\right)\right]\right\} ^{2}. 
\end{multline}
As an example, in Fig.(\ref{fig:n_j}) we present the asymptotic time average starting from the
states $\ket{\Phi_0} = \ket{...01110...}, \ket{...0111110...}$ and $ \ket{...0101010...}$ for
different values of the applied tilt $E$.  Note that these quantities are the same for the cases
$U=0$, $U\to\infty$ and $V=0$.  (This solution does not apply to $V=\infty$, as explained in Section
\ref{suppsec_freepoints}.)

Note that here $\bar n_j$ and $\bar \sigma_j$ were computed performing the integral over time
explicitly.  Away from the integrable points, the same quantities given in the main text were
obtained using the diagonal ensamble expression: $\bar n_j = \sum_\alpha \abs{c_\alpha}^2
\bra{\alpha} n_ j \ket{\alpha} $ and $\bar \sigma_j^2 = \sum_{\alpha \neq \alpha'} \abs{c_\alpha}^2
\abs{c_{\alpha'}}^2 \abs{\bra{\alpha} n_ j \ket{\alpha'}}^2 $. Diagonal ensemble values are only
valid for a system with non-degenerate energy levels which is the case for finite
interactions. Therefore the limits $U,V\to 0, \infty$ of the quantities obtained in the main text do
not coincide with those computed here.  One way of understanding this is that, in the computation of
the asymptotic long time averages, the limits $t\to\infty$ and $U,V\to 0, \infty$ do not commute.

\section{Beating vs equilibration behaviors  \label{suppsec_beating}}

In this section we expand on the results reported in the main text concerning many-body Bloch
dynamics slightly away from the non-interacting  points, i.e., the Ft-VM at $V\ll J$ and $V\gg J$, and
the BHM at $U\ll J$ and $U\gg J$.

\subsection{Spectral explanation of beating frequency}

%%%%%%%%%%%%% FIGURE IN SUPPLEMENTARY %%%%%%%%%%%%%%%%%%%%%%%%%%%%%%%
\begin{figure}
\begin{centering}
\includegraphics[width=\columnwidth]{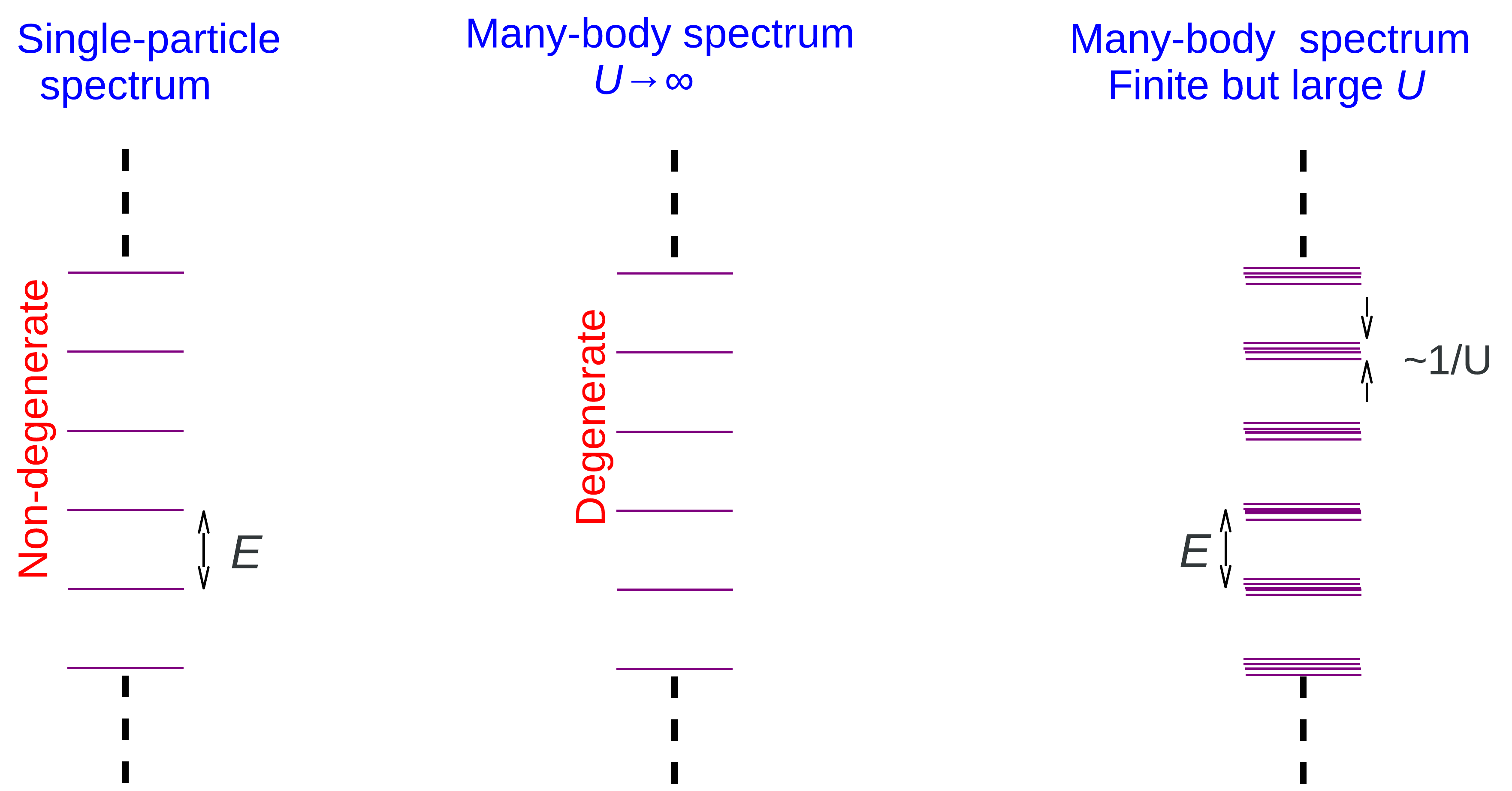}
\end{centering}
\caption{\label{fig_spectrum_cartoon}
Spectral explanation of beating frequency scaling as $\sim U^{-1}$ for large $U$ in the Bose-Hubbard
model.  (The same explanation holds for the $\sim V^{-1}$ behavior of the fermionic $t$-$V$ model
and the $\sim U$, $\sim V$ behaviors at small interactions.)
Left: The single-particle spectrum is non-degenerate and equally spaced.  Center: The many-body
spectrum for the non-interacting case is obtained by filling up the single-particle levels in all
possible ways, hence it is also equally spaced, but each many-body eigenenergy is massively
degenerate.  Right: moving away from the free point, the degeneracies get lifted, so that each
energy level is broadened.  In the perturbative regime, the broadening is $O(U^{-1})$.
}
\end{figure}
%%%%%%%%%%%%% FIGURE IN SUPPLEMENTARY %%%%%%%%%%%%%%%%%%%%%%%%%%%%%%%

For the BHM, the beating frequency is $\propto U^{-1}$ for $U\gg J$ and $\propto U$ for $U \ll J$, as
shown in Figure 2 of the main text by plotting the beat period against $U$.  For the Ft-VM, the
beating frequency will similarly be $\propto V^{-1}$ for $V\gg J$ and $\propto V$ for $V\ll J$, near the
two non-interacting points.  

As announced in the main text, a perturbative argument starting from the corresponding 'free' point
($U,V=0,\infty$) explains this behavior.  This is illustrated in Figure \ref{fig_spectrum_cartoon}
(and expanded below) for the case of large $U$.  Exactly the same argument holds for small $U$ and
for the Ft-VM.  One simply has to replace $1/U$ by the corresponding perturbative parameter ($U$,
$V$ or $1/V$), as appropriate.

%%%%%%%%%%%%% FIGURE IN SUPPLEMENTARY %%%%%%%%%%%%%%%%%%%%%%%%%%%%%%%
\begin{figure}
\centering
\includegraphics[width=1\columnwidth]{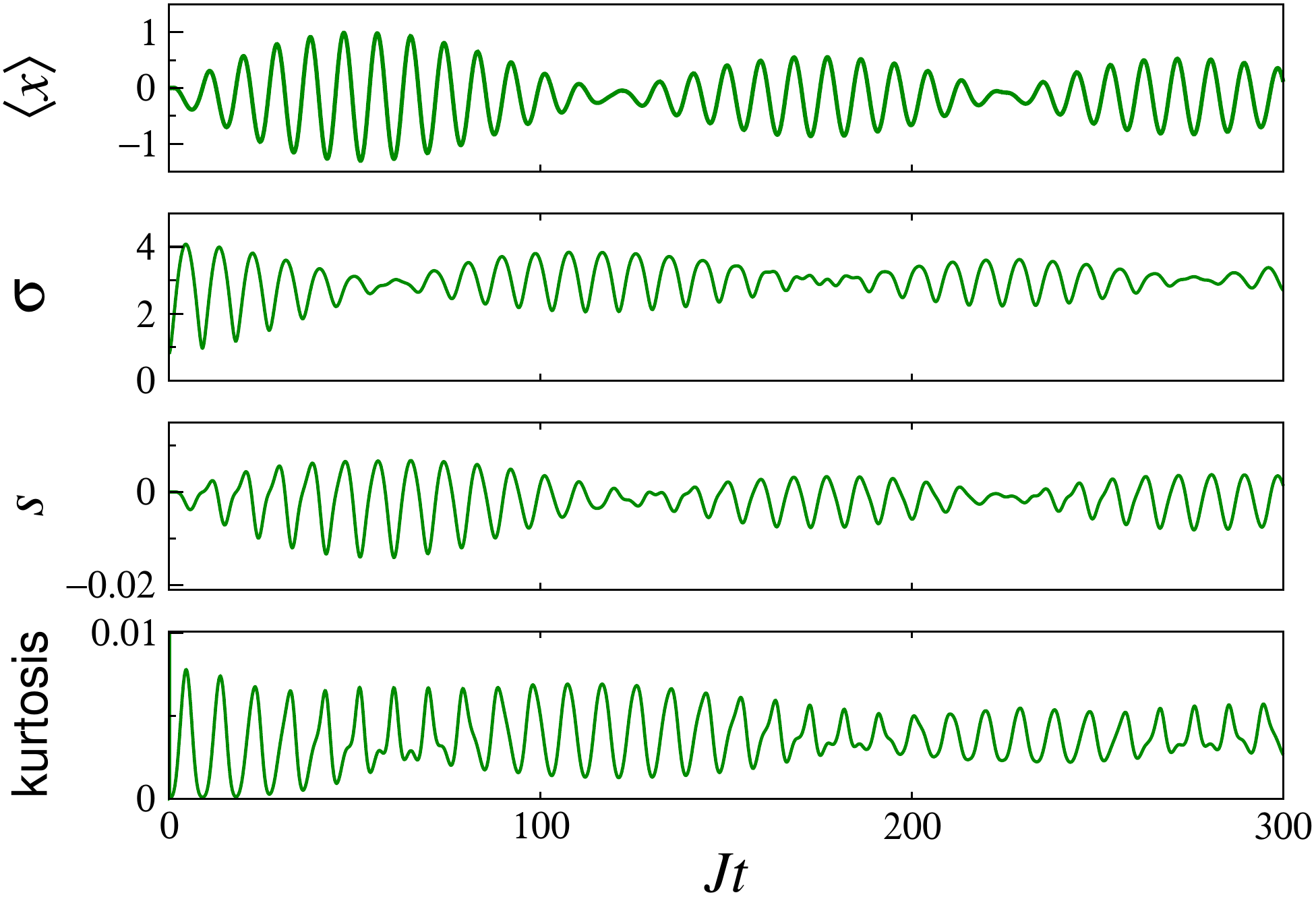}
\caption{\label{fig:skewness_kurtosis}
Dynamics of three bosons starting at an initial product-state configuration $...0011100...$, with  $E=0.35$
and $U=20$.  The dynamics of the cloud is shown through the time evolution of first four moments of the
occupancy distribution: the center of mass $<x>$, the r.m.s.\ width $\sigma$, the skewness $s$ and
the kurtosis.  Beating behavior with the same beating period is seen in all these quantities.  
}
\end{figure}
%%%%%%%%%%%%% FIGURE IN SUPPLEMENTARY %%%%%%%%%%%%%%%%%%%%%%%%%%%%%%%

The spectrum of a single particle in a Stark ladder has equally spaced non-degenerate levels with
spacing $E$.  Note that, each eigenstate corresponds to localization around a particular site. 
For a non-interacting many-body system, the many-body spectrum can be constructed out of the
single-particle spectrum by filling the single-particle levels with various numbers of particles.
In this case, the many-body eigenenergies are sums of single-particle eigenenergies.  
Hence the possible values of the many-body eigen-energies are also equally spaced with spacing $E$.  
However, these levels are now highly degenerate, as many different combinations of single-particle eigenstates can lead to the
same total eigenenergy.  For a fixed number $N$ of particles, the degeneracy of a level far from the
spectral edges would scale with the number of sites as $O(L^{N-1})$ (ignoring the fact that, for any
finite $L$, the levels are not exactly equally spaced and so the degeneracies are not exact).  In
the limit $L\to\infty$ that we are interested in, each level is infinitely degenerate for any
$N>1$, and the degeneracies are exact.  

Since the levels are all equally spaced with spacing $E$, the time evolution of any observable
quantity will be perfectly periodic with period $2\pi/E$.

Now we consider perturbing the system by moving away slightly from the `free' points.  The
perturbation parameter is $1/U$ for the BHM system near the hard-core limit.  The perturbation will
lift the degeneracies, and the splitting is proportional to the $1/U$.  The time evolution of any
observable quantity under this Hamiltonian can be written (when the initial state is expanded in the
energy eigenstate basis) as a sum of oscillating terms, with the oscillation frequencies being the
energy differences between eigenstates.  Now, because of the splitting as shown in Figure
\ref{fig_spectrum_cartoon} (right panel), the frequencies are not all equal to $E$, rather they are
clustered around the value $E$ with frequency difference of order $1/U$.  This explains why the beat
periods are inversely proportional to the perturbation parameter.

\subsection{Beating in different quantities, in both BHM and Ft-VM}

In the main text, we presented time evolution data displaying beating behavior in the center of mass
and width of the cloud, for the BHM, and reported that the same behavior can be seen in other
observables, such as the skewness and kurtosis of the cloud.  This is shown in Fig.
\ref{fig_spectrum_cartoon}.  The beating behavior is also visible in the time evolution of the site
occupancy or double occupancy (not shown).

%%%%%%%%%%%%% FIGURE IN SUPPLEMENTARY %%%%%%%%%%%%%%%%%%%%%%%%%%%%%%%
\begin{figure}
\begin{centering}
\includegraphics[width=0.75\columnwidth]{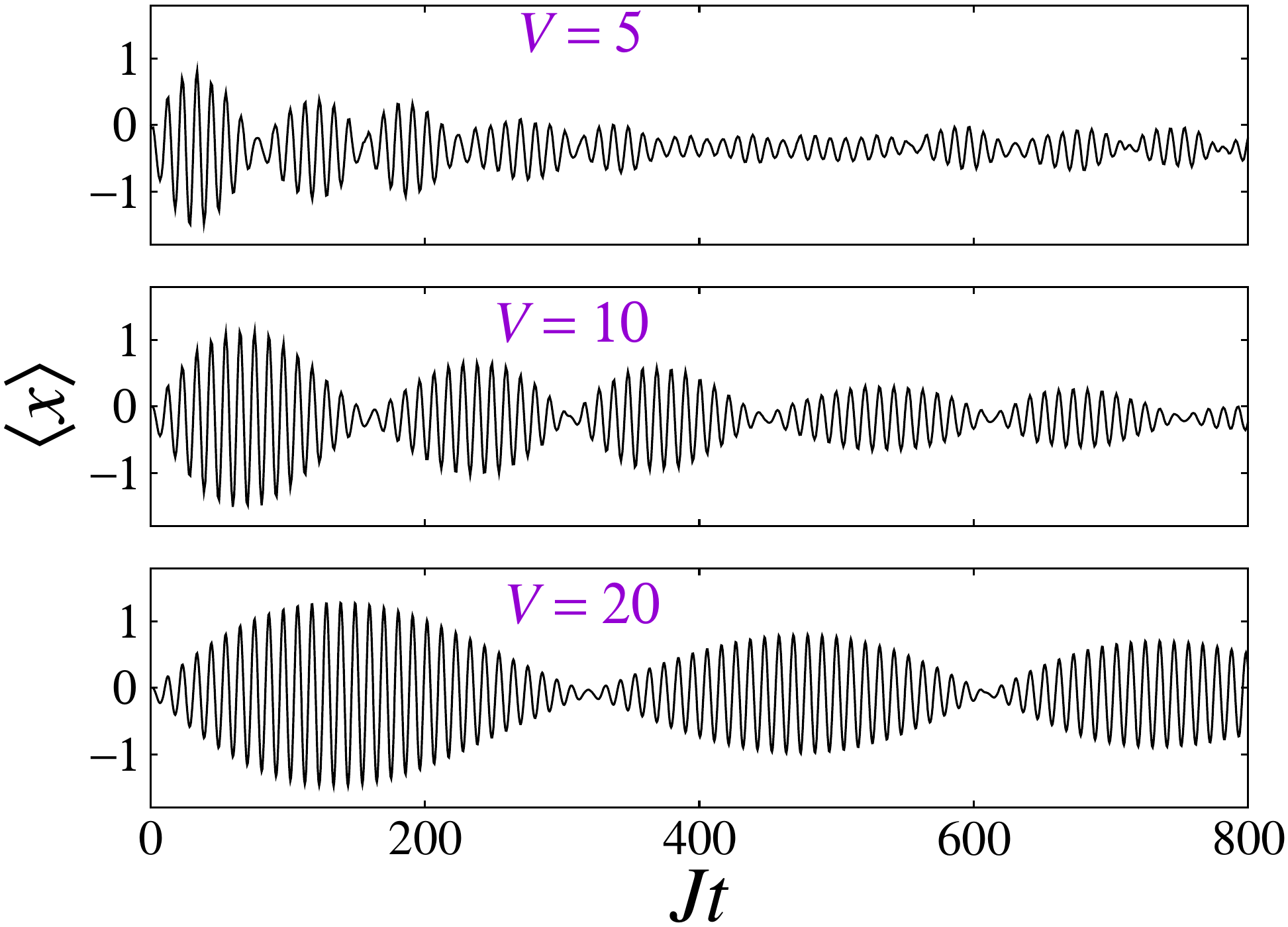}
\end{centering}
\caption{\label{fig:tV_large_varying_V} Dynamics of the cloud center of mass for the Ft-VM, with three
  fermions starting at an initial product-state configuration $...001010100...$.  The Stark field is
  $E=0.3$; three large values of the interaction strength $V$ are shown.  The beat frequency can be
  seen to vary as $V^{-1}$.  (Beat period doubles when $V$ is doubled.) 
}
\end{figure}
%%%%%%%%%%%%% FIGURE IN SUPPLEMENTARY %%%%%%%%%%%%%%%%%%%%%%%%%%%%%%%

In the main text, we also reported that the same beating behavior appeared in the fermionic system.
This is shown in Figure \ref{fig:tV_large_varying_V}.

%% %%%%%%%%%%%%% FIGURE IN SUPPLEMENTARY %%%%%%%%%%%%%%%%%%%%%%%%%%%%%%%
%% \begin{figure}
%% \begin{centering}
%% \includegraphics[width=1\columnwidth]{ppr_BH_box_initial_02}
%% \end{centering}
%% \caption{\label{fig:Finite_U}}
%% \end{figure}
%% %%%%%%%%%%%%% FIGURE IN SUPPLEMENTARY %%%%%%%%%%%%%%%%%%%%%%%%%%%%%%%

%% %%%%%%%%%%%%% FIGURE IN SUPPLEMENTARY %%%%%%%%%%%%%%%%%%%%%%%%%%%%%%%
%% \begin{figure}
%% \begin{centering}
%% \includegraphics[width=1\columnwidth]{site_OccupancyDynamics_varyU_A_01}\tabularnewline
%% \end{centering}

%% \caption{\label{Spectrum...}}
%% \end{figure}
%% %%%%%%%%%%%%% FIGURE IN SUPPLEMENTARY %%%%%%%%%%%%%%%%%%%%%%%%%%%%%%%

%%%%%%%%%%%%% FIGURE IN SUPPLEMENTARY %%%%%%%%%%%%%%%%%%%%%%%%%%%%%%%
\begin{figure}
\begin{centering}
\includegraphics[width=1\columnwidth]{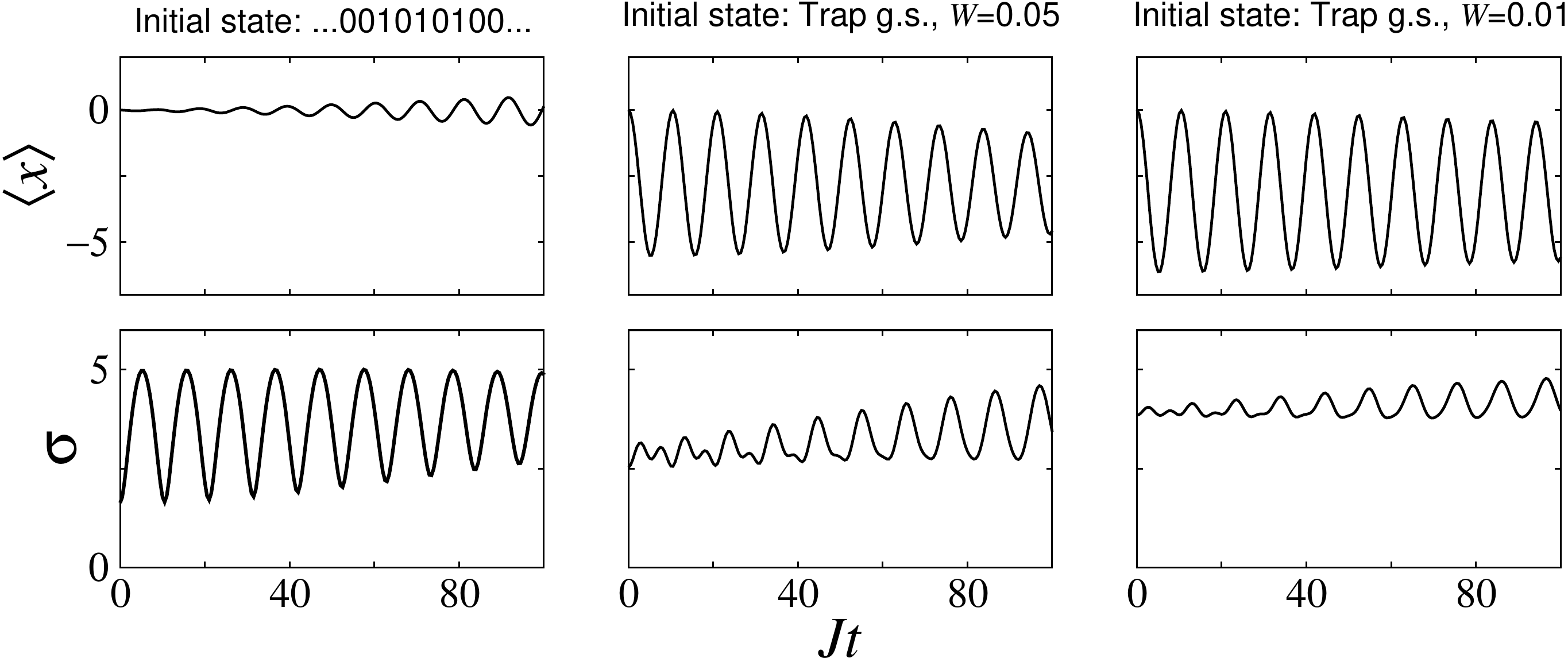}
\end{centering}
\caption{ \label{Ft-VM_vary_initial} Fermionic t-V model, dynamics of 3 fermions, $E=0.3$, $V=0.1$.
  The initial state is changed from left to right.  The product initial state (...001010100...) used
  for the leftmost panels is a proxy for box-like (tightly-trapped) initial state while the
  rightmost column corresponds to  weakest trapping.
}
\end{figure}
%%%%%%%%%%%%% FIGURE IN SUPPLEMENTARY %%%%%%%%%%%%%%%%%%%%%%%%%%%%%%%

\subsection{$\Delta{x}$ and $\Delta{\sigma}$ as function of trap strength}

For the non-interacting system ($V=0$ for Ft-VM and $U=\infty$ for BHM), we have shown that Bloch
oscillations are primarily position oscillations (large $\Delta{x}$, small $\Delta{\sigma}$) if the
trapping is weak and the initial cloud shape is gaussian-like, while the oscillations are
predominantly width oscillations (small $\Delta{x}$, large $\Delta{\sigma}$) if the trapping is
strong and the initial cloud is `box'-shaped.

%% Try getting data away from the perturbative regime (for $V=1$) as well.  

%% Also, maybe have three trap strengths rather than the 10101 state. 

In Figure \ref{Ft-VM_vary_initial}, we demonstrate that some traces of this phenomenon survive, at
least initially, when interactions are added.  The example shown is for the fermionic model (Ft-VM),
for a small interaction.  The dynamics of $\langle{x}\rangle$ and $\sigma$ are shown for three different
initial states.  On the left column, the initial state is $...001010100...$, which may be considered as the
analog of a box-like initial state ($\tilde{\rho}=\infty$) for fermions.  The center and right
columns correspond to finite trap ground states, with the rightmost column corresponding to weaker
traps.  

Going from the leftmost to rightmost columns, the amplitude of center-of-mass oscillations gets 
larger, while the amplitued of width oscillations becomes smaller.  Thus the intuition of which type
of oscillation (position vs width) dominates, which we have gained from the non-interacting systems,
continues to be valid for interacting systems.

\section{Viewing the Many-Body Spectrum   \label{suppsec_spectrum}}

%%%%%%%%%%%%% FIGURE IN SUPPLEMENTARY %%%%%%%%%%%%%%%%%%%%%%%%%%%%%%%
\begin{figure}[b]
\begin{centering}
\includegraphics[width=1\columnwidth]{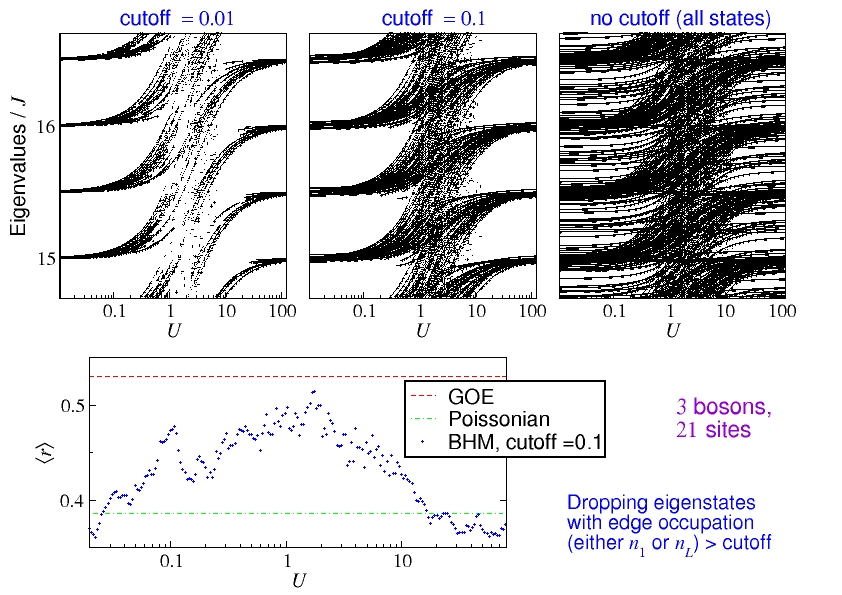}
\end{centering}
\caption{\label{supp_Spectrum} Top panels: spectrum of BHM, obtained numerically with $N_p=3$ bosons
  in $L=21$ sites, $E=0.25$.  In the left and center panel, any eigenstate with edge occupancy
  ($n_1$ or $n_L$) larger than the indicated cutoff is omitted.  Bottom panel: The average of ratios of
  consecutive level spacings in the spectrum obtained after the cutoff procedure.  
}
\end{figure}
%%%%%%%%%%%%% FIGURE IN SUPPLEMENTARY %%%%%%%%%%%%%%%%%%%%%%%%%%%%%%%

%% Discuss effect of boundaries, how can be `fixed' by removing states with heavy boundary weights. 

We have made several arguments about the many-body dynamics based on the many-body spectrum for
fixed $N_p$ and infinite $L$. 
Unfortunately, it is not possible to numerically calculate the many-body spectrum explicitly for
$L\to\infty$.  Nevertheless, we can still use the results of finite-size numerical diagonalizations
to infer relevant features of the many-body spectrum. 

One issue arising with finite-$L$ data is that, even for a single particle, the spectrum is not
exactly equally spaced.  The deviation is more severe near the edges of the spectrum, i.e., for
eigenstates with a significant occupancy near the edges of the finite lattice.  For eigenstates
localized far from the edges, the corresponding single-particle eigenvalues are nearly equally
spaced.

In the main text, we employed the trick of intensity-coding (color-coding) the many-body eigenstates
by the overlap with a many-particle state trapped near the center of the lattice.  This ensures that
the edges of the lattice plays no role, so that we obtain an $L$-independent picture.

In Figure \ref{supp_Spectrum}, we use a complementary procedure to provide another view of the
many-body spectrum.  The numerically calculated eigenstates are filtered so that any eigenstate with
a significant occupancy at one of the edge sites is not shown.  A view of the central part of the
spectrum is shown for various values of the cutoff.  It seems reasonable to presume that, for the
infinite chain, the many-body spectrum in any slice of energy is similar to the picture obtained
with cutoff 0.1, with a more dense spectrum at intermediate $U$.  

In the bottom panel, we have analyzed the level statistics of the spectrum obtained for $L=21$ with
cut-off 0.1.  For intermediate $U$, the value approaches that expected for a chaotic system (GOE
value).  In the infinite-size limit, it is expected that the spectrum is still not chaotic and will
be Poissonian at all $U$ due to many-body localization.  (Eigenstates localized in different regions
can be expected not to interact with each other; hence have no level repulsion.)  However, our
cutoff procedure is biased toward eigenstates trapped near the center of the lattice, and we have
shown that if one restricts to the part of the Hilbert space localized or trapped in a particular
region then these systems behave like thermalizing (ETH-obeying or chaotic) systems.  This is
visible in our analysis of level statistics in the bottom panel of Figure  \ref{supp_Spectrum}.

\end{document}